\documentclass[11pt,a4paper]{article}
\pdfoutput=1

\usepackage[usenames,dvipsnames,svgnames,table,x11names]{xcolor}
\usepackage{graphicx}
\usepackage{bm}
\usepackage{amsmath,amsfonts,amssymb}

\usepackage[colorlinks,pdfstartview=FitH]{hyperref}
\hypersetup{linkcolor=blue,citecolor=blue,filecolor=black,urlcolor=blue}

\usepackage{cite}

\DeclareMathOperator{\im}{Im}

\usepackage[margin = 1in]{geometry}

\newcommand\LY{{\rm LY}}
\newcommand\WF{{\rm WF}}
\renewcommand\O{{\rm O}}
\newcommand\ord{\mathcal{O}}
\renewcommand\c{{\rm c}}
\renewcommand\sp{{\rm sp}}
\renewcommand\epsilon{\varepsilon}
\newcommand\giv{u_0}
\newcommand\givt{\widetilde{u}_0}
\newcommand\gf{{u}}
\newcommand\givf{\giv^\WF}
\newcommand\giii{g_3}
\newcommand\giiit{\widetilde{g}_3}

\makeatletter
\@addtoreset{equation}{section}
\makeatother

\title{On spinodal points and Lee-Yang edge singularities}

\author{X.\ An${}^{\textrm{a}}$, D.\ Mesterh\'azy${}^{\textrm{b}}$, and M.\ A.\ Stephanov${}^{\textrm{a}}$ \\[6pt]
  {\small ${}^{\textrm{a}}$Department of Physics, University of Illinois at Chicago} \\
  {\small 845 West Taylor Street, Chicago, IL 60607, USA} \\[4pt]
  {\small ${}^{\textrm{b}}$Albert Einstein Center for Fundamental Physics} \\
  {\small Institute for Theoretical Physics, University of Bern} \\
  {\small Sidlerstrasse 5, 3012 Bern, Switzerland}}

\begin{document}

\maketitle

\begin{abstract}
  \normalsize
  We address a number of outstanding questions associated with the analytic properties of the universal equation of state of the $\phi^4$ theory, which describes the critical behavior of the Ising model and ubiquitous critical points of the liquid-gas type. We focus on the relation between spinodal points that limit the domain of metastability for temperatures below the critical temperature, i.e., $T < T_\c$, and Lee-Yang edge singularities that restrict the domain of analyticity around the point of zero magnetic field $H$ for $T > T_\c$. The extended analyticity conjecture (due to Fonseca and Zamolodchikov) posits that, for $T < T_\c$, the Lee-Yang edge singularities are the closest singularities to the real $H$ axis. This has interesting implications, in particular, that the spinodal singularities must lie {\it off} the real $H$ axis for $d < 4$, in contrast to the commonly known result of the mean-field approximation. We find that the parametric representation of the Ising equation of state obtained in the $\epsilon = 4-d$ expansion, as well as the equation of state of the $\O(N)$-symmetric $\phi^4$ theory at large $N$, are both nontrivially consistent with the conjecture. We analyze the reason for the difficulty of addressing this issue using the $\epsilon$ expansion. It is related to the long-standing paradox associated with the fact that the vicinity of the Lee-Yang edge singularity is described by Fisher's $\phi^3$ theory, which remains nonperturbative even for $d\to 4$, where the equation of state of the $\phi^4$ theory is expected to approach the mean-field result. We resolve this paradox by deriving the Ginzburg criterion that determines the size of the region around the Lee-Yang edge singularity where mean-field theory no longer applies.
\end{abstract}

\newpage

\section{Introduction}
\label{sec:Introduction}

The universality of critical phenomena makes the knowledge of the equation of state of the Ising model or, more broadly, $\phi^4$ field theory, important to the study of a wide range of phenomena \cite{Pelissetto:2000ek} from Curie points in magnets and liquid-gas transitions, to the cosmologically relevant phase transition in the gauge-Higgs sector of the Standard model \cite{Rummukainen:1998as}, and the phase diagram of QCD at finite density studied in heavy-ion collisions \cite{Stephanov:1998dy,Stephanov:2004wx}. Owing to universality and scaling, the equation of state sufficiently close to the critical point, i.e., in the scaling region, can be characterized by a universal function of a single argument, a scale-invariant combination of two relevant variables -- the magnetic field and the temperature. Determining this function is a well-posed mathematical problem which to this day, however, remains unsolved, at least in the analytically exact sense. Nevertheless, a lot is known about the equation of state \cite{Pelissetto:2000ek}. This includes celebrated exact results, such as the Onsager solution of the two-dimensional Ising model in the absence of the magnetic field \cite{Onsager:1943jn} or the Lee-Yang theorem regarding the distribution of zeros of the partition function in the complex plane of the magnetic field variable \cite{Yang:1952be,Lee:1952ig}. The equation of state near the upper critical dimension, $d = 4$, is also understood in terms of the perturbative Wilson-Fisher fixed point using the $\epsilon = 4-d$ expansion \cite{Brezin:1972fc,Avdeeva:1972,Wallace:1973,Wallace:1974,Nicoll:1985zz}. Furthermore, there are numerous numerical studies based on the high-temperature series expansion \cite{Campostrini:1999,Campostrini:2002}, perturbative field-theory expansions \cite{Guida:1996ep}, Monte Carlo lattice simulations \cite{Tsypin:1994by,Tsypin:1994nh,Tsypin:1997zz,Caselle:1997hs,Hasenbusch:1998gh}, the exact renormalization group \cite{Berges:1995mw}, as well as the truncated free-fermion space approach \cite{Fonseca:2001dc}.

In this paper we focus on the analytic properties of the universal equation of state in the scaling regime near the Ising critical point as a function of a {\it complex} magnetic field $H$. Two notable facts will guide our discussion. The first is Lee and Yang's observation that the singularities in the complex magnetic field plane terminate two (complex conjugate) branch cuts, which according to the Lee-Yang theorem \cite{Lee:1952ig}, must lie on the imaginary axis. These branch points, or Lee-Yang edge singularities, ``pinch'' the real axis as the temperature $T$ approaches its critical value $T_\c$ from above, resulting in a singularity on the real axis at zero magnetic field -- the Ising critical point. The second is the observation by Fisher that the thermodynamic singularity at the Lee-Yang edge point corresponds to the critical point in the $\phi^3$ theory \cite{Fisher:1978pf}. The upper critical dimension of this theory is six, which means that below this dimension the critical exponent $\sigma$ that characterizes the vanishing of the discontinuity at the Lee-Yang branch point is not simply given by its mean-field value $1/2$. This includes the case $d = 4-\epsilon$, where the Ising equation of state is believed to be described by mean-field theory with corrections suppressed by $\epsilon$. Here, we address the apparent contradiction between the conclusions of Fisher's analysis and the $\epsilon$ expansion around $d = 4$.

Analyticity of the equation of state allows one to connect high- and low-temperature domains near the critical point \cite{Privman:1982a,Penrose:1993}. In particular, using the mean-field equation of state one can show that the Lee-Yang edge singularities, which reside on the imaginary magnetic field axis, are analytically connected to singularities that limit the domain of metastability -- so-called spinodal singularities \cite{Fisher:1967dta,Binder:1974,Binder:1987}. The latter reside on another Riemann sheet reachable by analytic continuation through the branch cut along the real magnetic field axis, describing the first-order phase transition at zero magnetic field. The position of these singularities on the real axis, however, is an artifact of the mean-field approximation. In fact, in $4-\epsilon$ dimensions the position of the spinodal point shifts into the complex plane by an amount of order $\epsilon^2$. We analyze this phenomenon in the framework of the $\epsilon$ expansion employing parametric representations of the equation of state \cite{Schofield:1969,Josephson:1969,Schofield:1969zz,Schofield:1973}. Our goal is to confront the extended analyticity conjecture advanced by Fonseca and Zamolodchikov \cite{Fonseca:2001dc}, which states that the complexified spinodal point is the nearest singularity to the real axis of the magnetic field.

We point out that our analysis is not complete, since the $\epsilon$ expansion fails to capture certain nonperturbative aspects of the universal Ising equation of state, most notably the Langer cut \cite{Langer:1967ax}. However, as we shall see, other important questions can nevertheless be addressed within such an approach. One has to bear in mind also that our results apply to the scaling region where the universal behavior is observed. However, many of the conclusions, such as those pertaining to the Langer cut, associated metastability and the shift of the spinodal point into the complex $H$-plane due to fluctuations should, arguably, remain true outside the scaling region.

We hope that the insights our study provides will contribute to a more complete picture of the $\phi^4$ theory. In particular, our work could help develop better parametrizations of the equation of state by taking into account its correct analytic properties. The knowledge of the complex singularities of the equation of state is also important for determining the position of the QCD critical point using lattice Taylor expansion methods \cite{Stephanov:2006dn}.

The outline of this article is as follows: In Sec.\ \ref{sec:Critical equation of state and the mean-field approximation} we review the properties of the mean-field equation of state of the scalar $\phi^4$ theory, and introduce the Lee-Yang edge singularities with their low-temperature image -- the spinodal points. Next, in Sec.\ \ref{sec:Beyond the mean-field approximation}, we discuss the limitations of the mean-field approximation. In particular, we derive the Ginzburg criterion which quantifies the breakdown of mean-field theory near the Lee-Yang edge singularities. Thereafter, in Sec.\ \ref{sec:Singularities in the epsilon expansion}, we employ the $\epsilon = 4-d$ expansion and examine the nature of the complex-field singularities in the framework of parametric representations of the Ising equation of state. In Sec.\ \ref{sec:Singularities in the O(N)-symmetric phi^4 theory} we consider the same problem from the point of view of the $\O(N)$-symmetric $\phi^4$ theory in the large-$N$ limit. In the concluding section, Sec.\ \ref{sec:Conclusions}, we summarize our findings. We argue that they are consistent with the extended analyticity conjecture put forward by Fonseca and Zamolodchikov and discuss the difficulty of establishing the latter rigorously in the $\epsilon$ expansion.

\section{Critical equation of state and the mean-field approximation}
\label{sec:Critical equation of state and the mean-field approximation}

The scalar $\phi^4$ theory in $d$ dimensions can be defined by the Euclidean action (or, depending on the context, the Hamiltonian divided by temperature)
\begin{equation}
  \label{eq:S}
  \mathcal{S} = \int d^dx \left[\frac{1}{2} (\partial_\mu\phi)^2 + \frac{r_0}{2} {\phi}^2 + \frac{\giv}{4!}\phi^4 - h_0 \phi \right] .
\end{equation}
The expectation value of the field $\phi$, $\langle\phi\rangle$, can be found by differentiating the logarithm of the partition function (free energy) with respect to $h_0$. The relation between the expectation value $\langle\phi\rangle$ and the bare parameters $r_0$ and $h_0$ (and, generally, also $\giv$ as well as the ultraviolet cutoff) defines the equation of state.

More specifically, we are interested in the critical point of this theory, i.e., the point in the parameter space where the correlation length $\xi$, measured in units of the cutoff scale, diverges. This point can be reached at $h_0 = 0$, by tuning $r_0\to r_\c$ for any given $\giv$. In fact, below the upper critical dimension, i.e., $d < 4$, the effective coupling runs into an infrared (IR) fixed point, the Wilson-Fisher fixed point \cite{Wilson:1971dc} and, as a result, the dependence on the coupling $\giv$ and the cutoff disappears -- the equation of state becomes a relation between three variables: $\langle\phi\rangle$, $h_0$, and $r_0$.

The critical $\phi^4$ theory provides a universal description of critical phenomena in many physically different systems such as liquid-gas or binary fluid mixtures or spin systems such as uniaxial ferromagnets. For the latter, the parameter $h_0$ can be mapped onto the applied external magnetic field, i.e., $h_0\sim H$, while 
\begin{equation}
  \label{eq:t-vs-r}
  t\equiv r_0 - r_\c ,
\end{equation}
is proportional to the deviation of the temperature from the critical (Curie) point, i.e., $t\sim T-T_\c$. 

In terms of conveniently rescaled variables $\Phi = \sqrt{\giv/6}\,\phi$ and $H = \sqrt{\giv/6}\,h_0$, the action \mbox{Eq.\ \eqref{eq:S}} takes the form
\begin{equation}
  \label{eq:S2}
  \mathcal{S} = \frac{6}{\giv} \int d^dx \left[\frac{1}{2} \left(\partial_\mu \Phi\right)^2 + V(\Phi) \right] ,
\end{equation}
with the potential 
\begin{equation}
  \label{eq:V}
  V(\Phi) = \frac{r_0}{2} \Phi^2 + \frac{1}{4} \Phi^4 -H \Phi .
\end{equation}
It is clear from Eq.\ \eqref{eq:S2} that for small $\giv$ fluctuations are suppressed and the path integral defining the partition function of the theory can be evaluated in the saddle-point, or mean-field, approximation. In this approximation the expectation value of the field, $\langle\Phi\rangle = M$, is a coordinate-independent constant that minimizes the potential \eqref{eq:V}, i.e.,
\begin{equation}
  \label{eq:MFEOS}
  V'(M) = -H + r_0 M + M^3 = 0 . 
\end{equation}

The correlation length $\xi$ is defined in terms of the second derivative of the (effective) potential $V$ at its minimum and, in the mean-field case, it is given by
\begin{equation}
  \label{eq:xi}
  \xi^{-2} = V''(M) = r_0 + 3 M^2 .
\end{equation} 
The Ising critical point, $\xi\to\infty$, is reached at $H = M = r_0 = 0$, and therefore
\begin{equation}
  \label{eq:tr0}
  t = r_0 .
\end{equation}
The implicit (multivalued) function $M(t,H)$ defined by Eq.\ \eqref{eq:MFEOS} represents the mean-field equation of state of the $\phi^4$ theory (or Ising model). 

\begin{figure}[!t] 
  \begin{picture}(0,120)
    \put(0,0){\includegraphics[width = 0.45\textwidth]{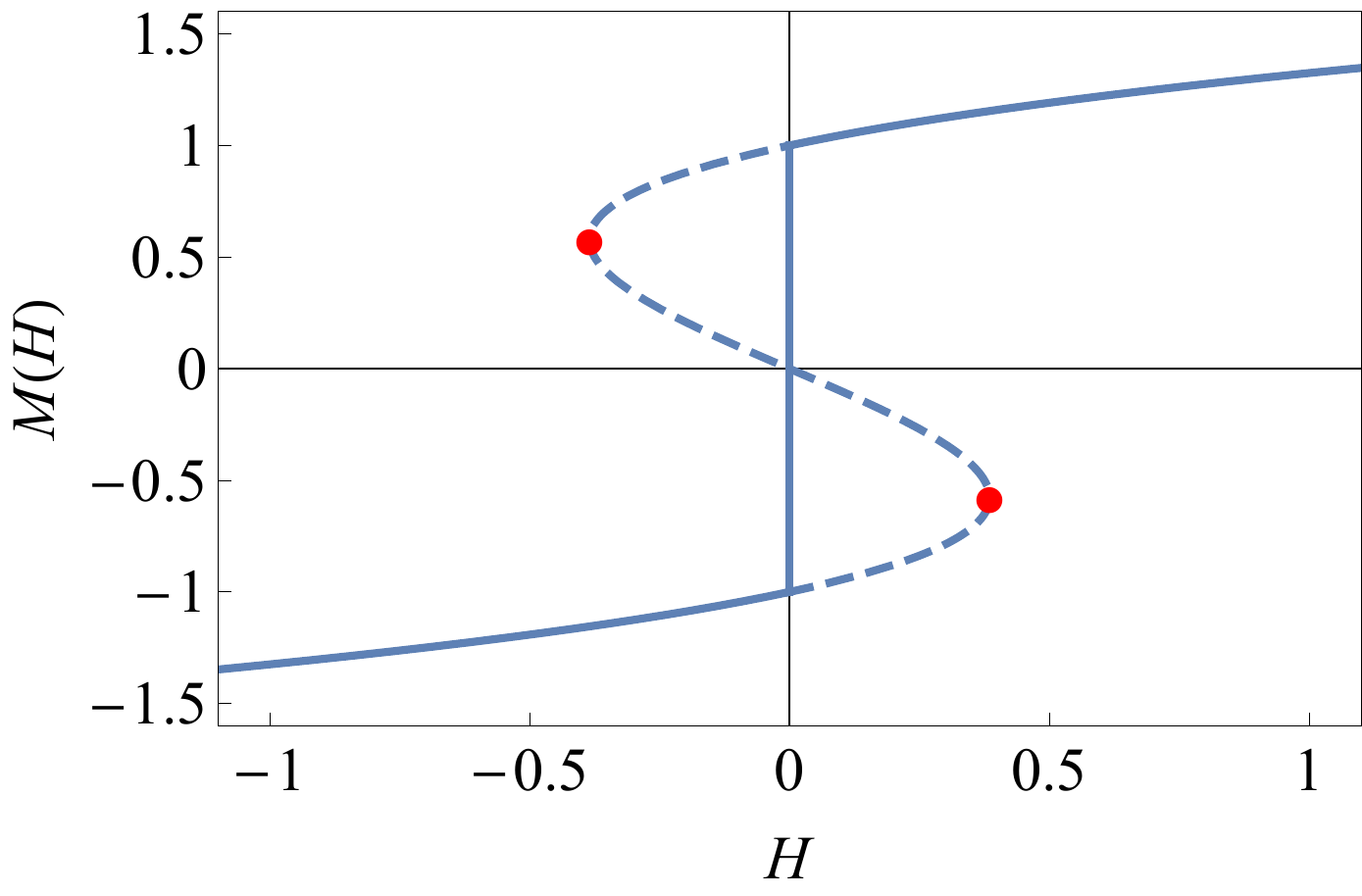}}
    \put(235,0){\includegraphics[width = 0.45\textwidth]{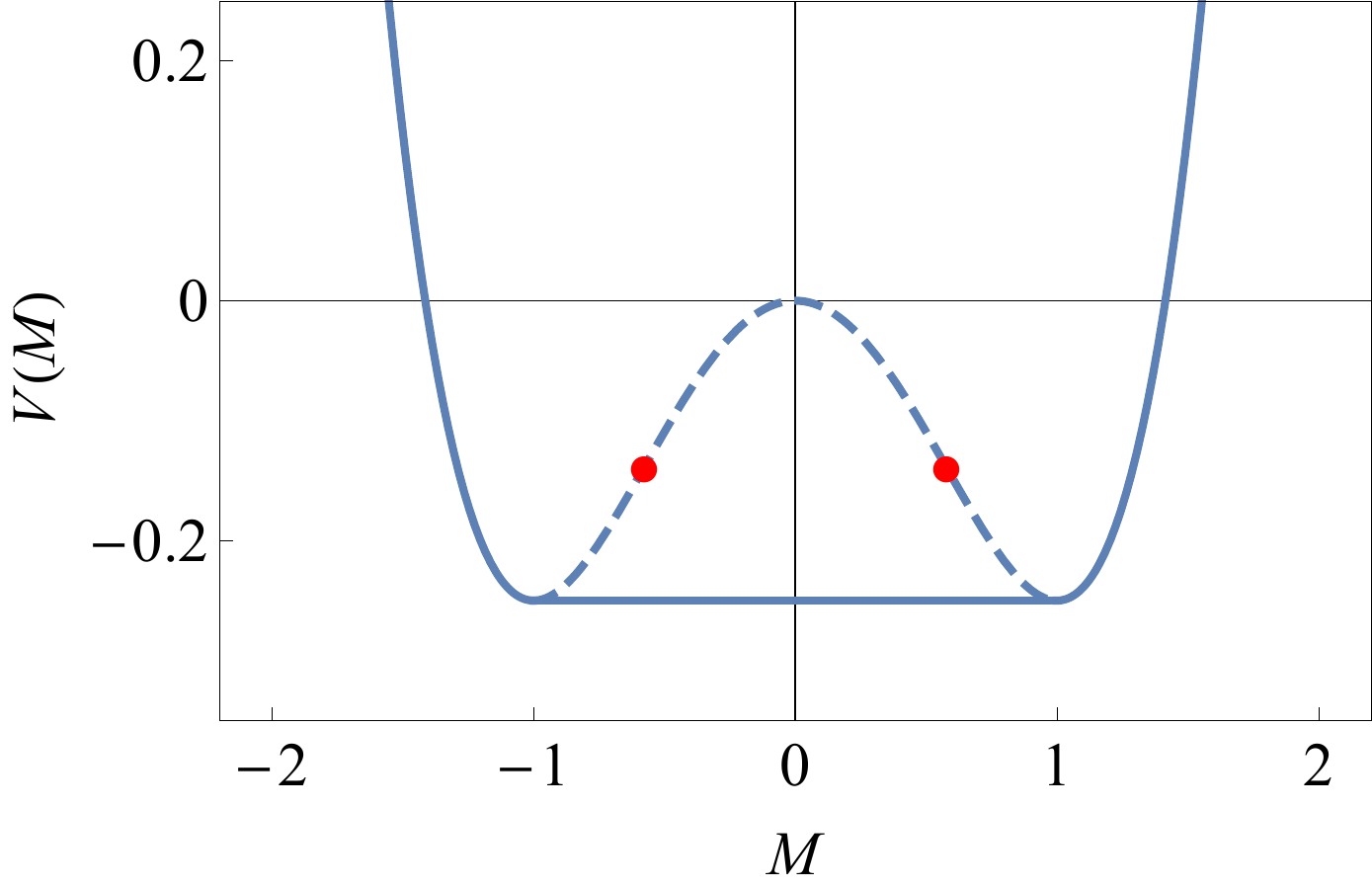}}
    \put(39,122){(a)}
    \put(274,122){(b)}
  \end{picture}
  \caption{\label{fig:1}(a) The mean-field Ising equation of state $M(H)$ and (b) the corresponding effective potential $V(M)$ at $H = 0$ in the low-temperature phase ($T < T_\c$). The analytic continuation of the stable branch (dashed curve) is bounded by the spinodal points (red). The straight line connecting the two minima of the effective potential is determined by the Maxwell construction.}
\end{figure}

It is clear from Eq.\ \eqref{eq:xi} that above the critical temperature of the Ising model, i.e., for $t = r_0 > 0$, the correlation length is finite for all {\it real} values of $H$. However, solving for $V'(M) = V''(M) = 0$, we find points on the {\it imaginary} axis, where $\xi\to\infty$ for $t > 0$:
\begin{equation}
  \label{eq:LY_MF}
  M_\LY = \pm \frac{1}{\sqrt{3}}\, i t^{1/2} \quad {\rm and} \quad
  H_\LY = \pm \frac{2}{3\sqrt{3}}\, i t^{3/2} .
\end{equation}
For $t > 0$, these branch points of $M(H)$, known as Lee-Yang (LY) edge singularities, terminate cuts that lie on the imaginary $H$ axis (according to the Lee-Yang theorem \cite{Yang:1952be,Lee:1952ig}). They pinch the real $H$ axis as the temperature $T$ approaches its critical value $T_\c$, i.e., $t\to 0$. 

On the other hand, below the critical temperature, $t < 0$, the mean-field approximation predicts that the correlation length, given by Eq.\ \eqref{eq:xi}, diverges at {\it real} values of $M$ and $H$. These so-called spinodal points are located on the metastable branch and limit the domain of metastability \cite{Fisher:1967dta,Binder:1974,Binder:1987}, as shown in Fig.\ \ref{fig:1}. 

\begin{figure}[!t]
  \centering
  \includegraphics[width = 0.61\textwidth]{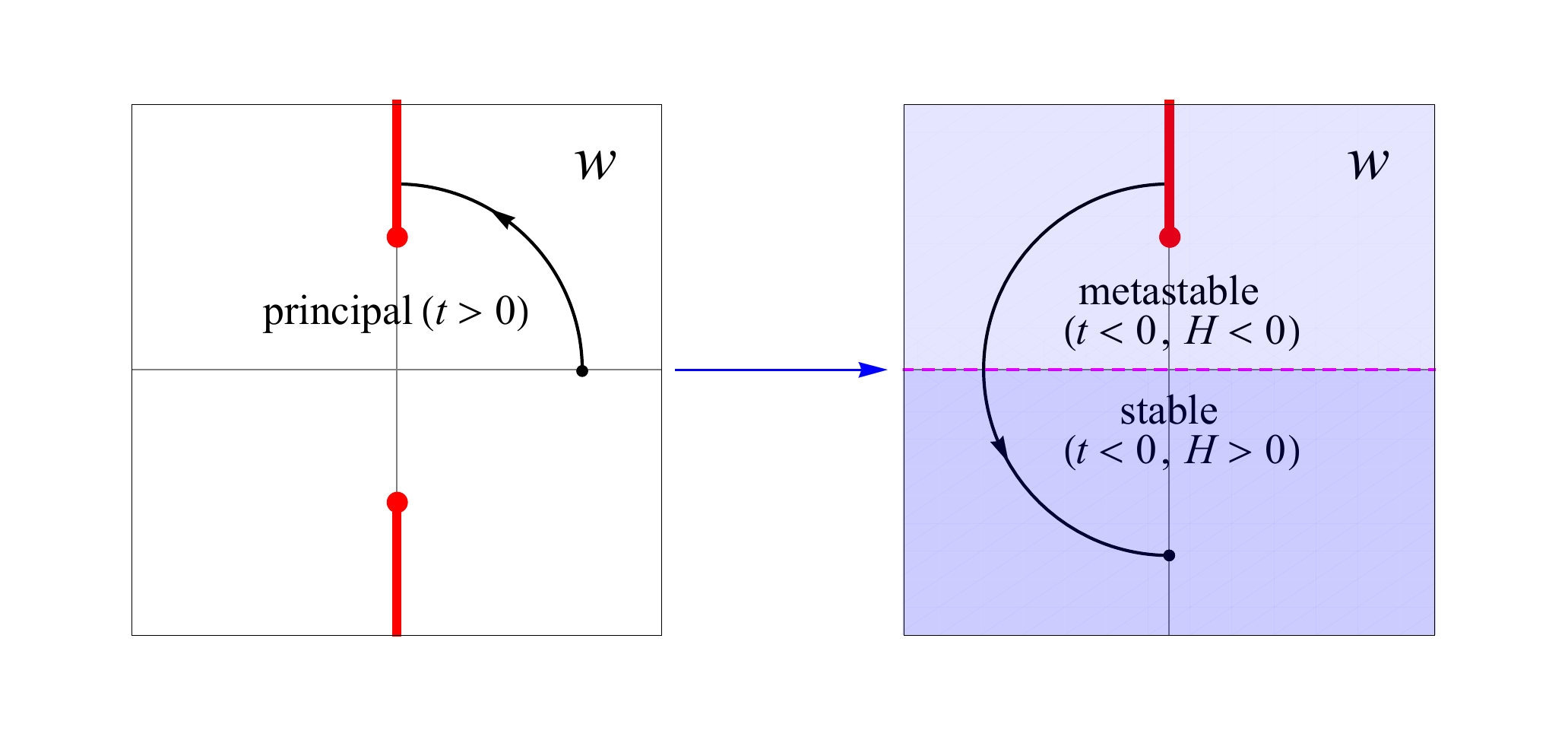}
  \caption{\label{fig:2}Analytic continuation $t\rightarrow -t$ from the principal, i.e., high-temperature sheet (left panel) to the low-temperature sheet (right panel) of the mean-field scaling function $z(w)$ in Eq.\ \eqref{eq:w=F(z)_MF} with $w \sim H t^{-3/2}$. Starting from $H > 0$ and $t > 0$, keeping $H > 0$ and $|t|$ fixed we rotate the phase $\arg t$ from 0 to $-\pi$ and trace the corresponding movement of the variable $w$ along the shown circular path. The principal sheet features a pair of Lee-Yang branch cuts along the imaginary $w$ axis, which terminate in the Lee-Yang edge singularities. Going through the cut we enter the metastable low-temperature branch ($H < 0$, $t < 0$). One reaches the stable branch ($H > 0$, $t < 0$) when $\arg t = -\pi$. From there one can also reach metastable branch $H < 0$ by rotating $\arg H$ from $0$ to $\pm\pi$, which changes $\arg w$ by $\pm\pi$.}
\end{figure}

An important property of the critical equation of state is scaling\footnote{Generally, scaling is a consequence of the coupling $\giv$ running into an IR fixed point.} \cite{Widom:1965,Fisher:1967dta}: The relation between $M$, $t$, and $H$ is invariant under simultaneous rescaling of these variables according to their scaling dimensions
\begin{equation}
  \label{eq:scaling}
  t\to \lambda t,\quad H\to \lambda^{\beta\delta} H, \quad {\rm and} \quad
  M\to\lambda^{\beta} M ,
\end{equation}
where $\beta$ and $\delta$ are standard critical exponents. The mean-field equation of state in Eqs.\ \eqref{eq:MFEOS}, \eqref{eq:tr0} scales with exponents
\begin{align}
  \label{eq:exponents_MF}
  \beta = \frac{1}{2} \quad {\rm and} \quad \delta = 3 \qquad \mbox{(mean field)}.
\end{align}

Scaling implies that the equation of state can be expressed as a relation between only two scaling-invariant variables. Depending on the choice of these variables it may be represented in several different ways. For example, we may express the equation of state in the Widom scaling form \cite{Widom:1965} 
\begin{equation}
  \label{eq:xy}
  y = f(x), \quad \mbox{with}\quad x \sim t M^{-1/\beta} \quad\mbox{and}\quad y \sim H M^{-\delta} ,
\end{equation}
where symbols `$\sim$' reflect arbitrary normalization constants which can be chosen to bring the function $f(x)$ into canonical form. Here, we express the mean-field scaling function $f(x)$ as
\begin{equation}
  \label{eq:y=f(x)_MF}
  f(x) = 1 + x ,
\end{equation}
with the scaling-invariant variables defined as
\begin{equation}
  x = t M^{-1/\beta}\quad\textrm{and} \quad y = H M^{-\delta} .
\end{equation}

However, the analytic properties as a function of $H$ at fixed $t$ are more manifest in another representation of the scaling equation of state
\begin{equation}
  \label{eq:wz}
  w = F(z),\quad\mbox{with}\quad w \sim H t^{-\beta\delta} \quad {\rm and} \quad z \sim M t^{-\beta} .
\end{equation}
Again, the normalization constants in Eqs.\ \eqref{eq:wz} can be chosen to achieve a conventional (canonical) form for the equation of state. We choose to express the mean-field scaling function $F(z)$ in the following form
\begin{equation}
  \label{eq:w=F(z)_MF}
  F(z) = z (1+z^2) ,
\end{equation}
with the variables
\begin{equation}
  \label{eq:sv_wz}
  w = H t^{-\beta\delta} \quad\textrm{and} \quad z = M t^{-\beta} .
\end{equation}

The inverse of the (mean-field) function $F(z)$, i.e., $z(w)$, is multivalued and has three Riemann sheets associated with the high- and low-temperature regimes of the mean-field equation of state. The principal sheet, which represents the equation of state $M(H)$ for $t > 0$, features two branch points. They are located on the imaginary axis in the complex $w$ plane
\begin{equation}
  \label{eq:zw_LY}
  w_\LY = \pm \frac{2i}{3\sqrt{3}} ,
\end{equation}
and correspond to the Lee-Yang edge singularities at imaginary values of the magnetic field $H$, cf.\ Eqs.\ \eqref{eq:LY_MF}. 

Going under either one of the associated branch cuts, e.g., by following the path shown in Fig.\ \ref{fig:2}, one arrives on the secondary sheet, which corresponds to the metastable branch of the equation of state at $t < 0$. The same branch point in Eq.\ \eqref{eq:zw_LY} viewed from this sheet represents the spinodal point located at real negative $H$. To arrive on the stable $t < 0$ branch, i.e., $H > 0$, one has to follow the circular path further in the anticlockwise direction, as shown in Fig.\ \ref{fig:2} (right). We conclude that, in the mean-field approximation, the spinodal points and the Lee-Yang edge singularities are manifestations of the {\it same} singularities of the scaling equation of state $z(w)$.

\section{Beyond the mean-field approximation}
\label{sec:Beyond the mean-field approximation}

The mean-field approximation relies on the smallness of the coupling $\giv$. This is justifiable for $d\ge 4$, where the coupling runs into the Gaussian IR fixed point and becomes arbitrary small as $\xi\to\infty$. For $d < 4$, the Wilson-Fisher (WF) fixed-point value of the coupling, $\givf = \ord(\epsilon)$, is also small as long as $\epsilon = 4-d\ll 1$. However, for the most interesting case $d = 3$ the theory is nonperturbative and we cannot rely on the mean-field approximation. We would like to address the following question: What happens with the spinodal points and Lee-Yang edge singularities in this case? We shall begin with general considerations and later consider the case of small $\epsilon$.

\subsection{Langer cut and Fonseca-Zamolodchikov conjecture}
\label{sec:Langer cut and Fonseca-Zamolodchikov conjecture}

According to the Lee-Yang theorem \cite{Yang:1952be,Lee:1952ig} the singularities of the Ising model, and thus, by universality, of the $\phi^4$ theory, must be located on the imaginary axis of $H$. Thus the result of the mean-field theory that the Lee-Yang edge singularities (and their associated cuts) are on the imaginary axis holds in general.\footnote{The theorem applies to singularities on physical stable branches of the function $M(H)$ (both below and above critical temperature) and thus cannot constrain the position of the spinodal singularities which are located on the metastable branch.}

What happens to the spinodal singularities away from mean-field? As we discussed, the scaling equation of state $z(w)$ describes both high- and low-temperature branches of $M(H)$, which correspond to primary and secondary Riemann sheets of the variable $w$. The Lee-Yang edge singularities are described by $w_\LY$, which lie on the imaginary $w$ axis because $w\sim Ht^{-\beta\delta}$ and for $t > 0$ the value of $H$ at the singularity, $H_\LY$, is imaginary. Thus analyticity and scaling imply that there must also be singularities on the low-temperature branch $t < 0$, at values of $H$ given by:
\begin{equation}
  \label{eq:H_spinodal_a.c.}
  H_\sp\sim w_\LY \hspace{1pt} t^{\beta\delta} = \mp |w_\LY \hspace{1pt} t^{\beta\delta}|\hspace{.5pt} e^{\pm i\pi(\beta\delta-3/2)} , \quad t < 0 ,
\end{equation}
where the different signs correspond to the two (complex conjugate) values of $w_\LY$ and to the two possible directions of rotation from $t$ to $-t = e^{\pm i\pi}t$. Thus, in general, the spinodal points $H_\sp$ are displaced from the (negative) real $H$ axis by a phase
\begin{equation}
  \label{eq:beta-delta-32}
  \Delta\phi = \pi\left(\beta\delta - \frac{3}{2}\right) ,
\end{equation}
where $\beta\delta > 3/2$ below the upper critical dimension, i.e., for $d < 4$ (cf.\ \mbox{Eq.\ \eqref{eq:Delta_epsilon}}), and $\beta\delta = 3/2$ for $d\geq 4$.

In order to understand the position of the points described by Eq.\ \eqref{eq:H_spinodal_a.c.} it is important to take into account another property of the equation of state in the low-temperature domain -- the Langer cut \cite{Langer:1967ax}. It is well-known that the Ising equation of state is weakly singular at $H = 0$ for $t < 0$, due to the presence of an essential singularity \cite{Andreev:1964,Fisher:1967dta,Isakov:1984we} associated with the decay of the metastable vacuum \cite{Kobzarev:1974cp,Coleman:1977,Callan:1977pt}. The rate of this decay gives the imaginary part of the free energy $\mathcal{F}(t,H)$ for $H$ on the metastable branch at $t < 0$ and, since $M = \partial\mathcal{F}/\partial H$, also the imaginary part of the magnetization $M(t,H)$. Near $d = 4$, it takes the form (for $w\ll1$)
\begin{equation}
  \label{eq:LangerSingularity}
  \im M(t, H) \sim \exp \left(-\frac{\rm const}{ \giv |w|^3}\right) ,
\end{equation}
demonstrating that there is an essential singularity, which is nonperturbative in $\giv$. Not only is this singularity absent in the mean-field equation of state, but it cannot be seen at any finite order of the $\epsilon$ expansion. The imaginary part of $M$ is discontinuous (changes sign by Schwarz reflection principle) across the real axis of $H$ on the metastable branch, which corresponds to a cut, known as the Langer cut \cite{Langer:1967ax}.

\begin{figure}[!t]
  \centering
  \includegraphics[width = 0.9\textwidth]{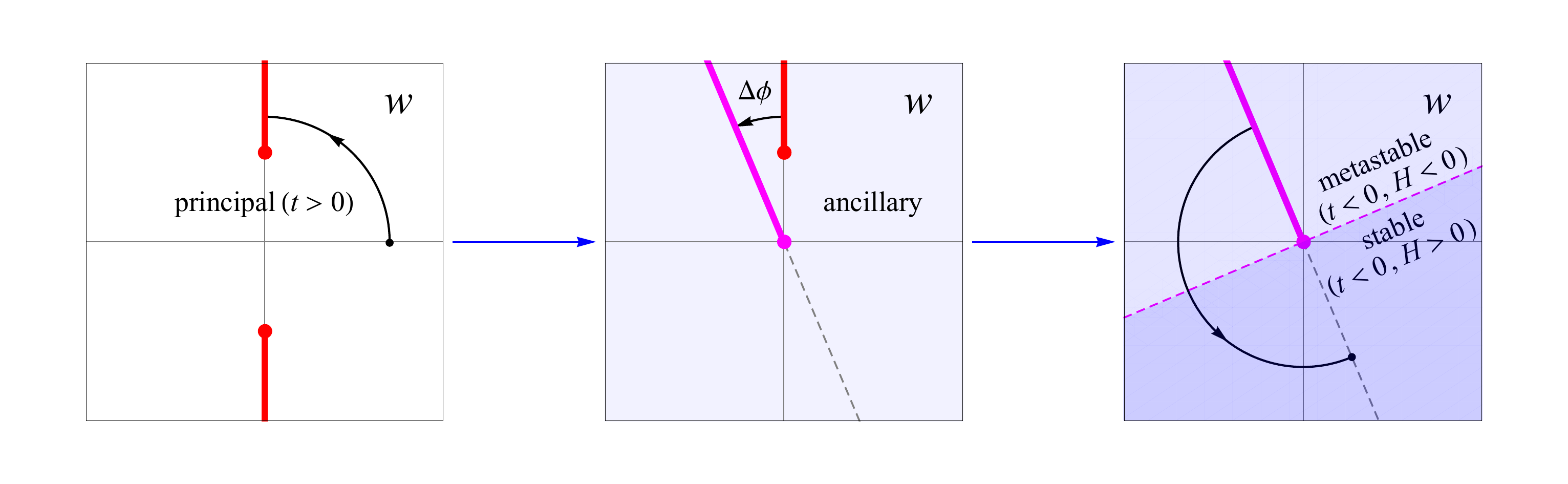}
  \caption{\label{fig:3}Analytic continuation $t\to -t $ from the principal, i.e., high-temperature sheet (left panel) to the low-temperature sheet (right panel) of the scaling function $z(w)$ of the Ising theory as conjectured by Fonseca and Zamolodchikov, where $w \sim H t^{-\beta\delta}$, while keeping the magnetic field $H > 0$ fixed at $d = 4-\epsilon$. After analytic continuation the metastable branch $H < 0$ can be accessed by rotating $H$ clockwise in the complex plane, while keeping $t < 0$ fixed. The line representing the Langer cut is rotated away from imaginary axis by an angle $\Delta\phi$, cf.\ Eq.\ \eqref{eq:beta-delta-32}.}
 \end{figure}

\begin{figure}[!t]
  \centering
  \includegraphics[width = 0.61\textwidth]{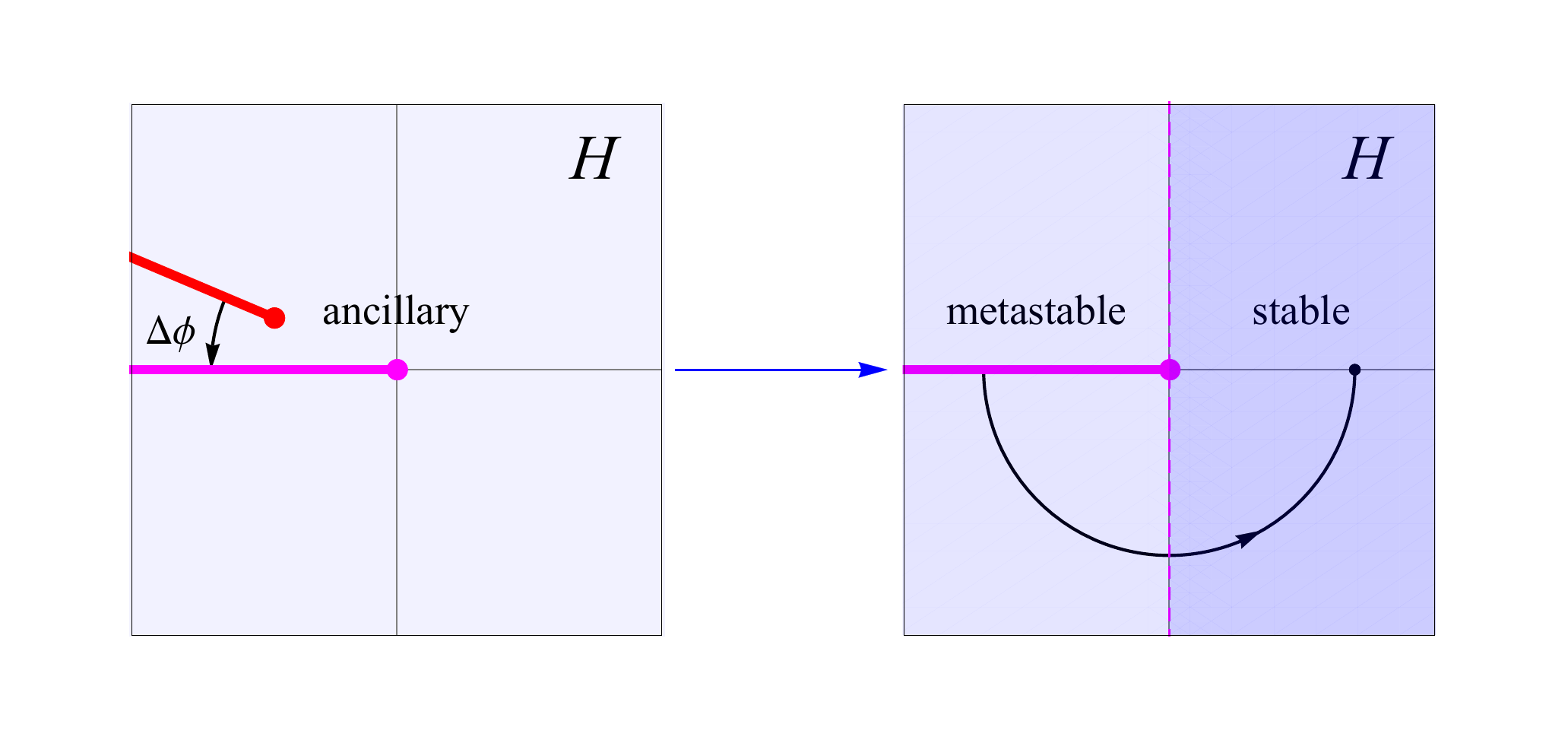}
  \caption{\label{fig:4}The Fonseca-Zamolodchikov conjecture for $t < 0$, illustrated in the complex $H$ plane. The line along the negative real $H$ axis represents the Langer cut. The second cut on the ancillary sheet is the Lee-Yang cut, which is associated with the Lee-Yang edge singularity. The latter is expected to be the nearest singularity under the Langer cut.} 
\end{figure}

This cut can be reached from the stable low-temperature branch ($H > 0$, $t < 0$) by rotating $H$ along a semicircle in the complex $H$ plane, such that $H\rightarrow -H$. Thus, its location in the complex $w$ plane should be as shown in Fig.\ \ref{fig:3}. If we translate Fig.\ \ref{fig:3} into the $H$ plane, using $w\sim H t^{-\beta\delta}$ (with $t < 0$), we find that the spinodal point can be found under the Langer cut as shown in Fig.\ \ref{fig:4}, assuming, of course, that we start from the stable $H > 0$ branch. It is therefore natural to expect that the spinodal singularity (which is {\it also} the Lee-Yang edge singularity) is the closest singularity to the real axis (i.e., to the Langer cut). This is the essence of the ``extended analyticity'' conjecture put forward by Fonseca and Zamolodchikov \cite{Fonseca:2001dc}. Here, our goal is to see what one can say about the singularities of the equation of state and the validity of the conjecture using the $\epsilon$ expansion as well as large-$N$ limit of the $\O(N)$-symmetric $\phi^4$ theory.

\subsection{Lee-Yang edge singularities and Ginzburg criterion}
\label{sec:Lee-Yang edge singularities and Ginzburg criterion}

As we discussed in Sec.\ \ref{sec:Critical equation of state and the mean-field approximation}, the mean-field (saddle-point) approximation is controlled by the quartic coupling $\giv$. For $d < 4$, in the scaling regime, the coupling is given by the IR (Wilson-Fisher) fixed-point value of order $\epsilon = 4-d$. This means that the true scaling equation of state should approach the mean-field one as $\epsilon\to0$. However, this approach is not uniform, especially, at the Lee-Yang edge singularities, which are the focus of this study.

The issue was first raised by Fisher, who observed that the singular behavior near the Lee-Yang point is described by a $\phi^3$ theory \cite{Fisher:1978pf}. This theory has an IR fixed point, albeit somewhat formally, since it occurs at {\it imaginary} values of the cubic coupling. The exponents (anomalous dimensions) can be calculated by an expansion around the upper critical dimension $d = 6$ of the $\phi^3$ theory where the theory becomes perturbative in terms of $\epsilon' = 6-d$. However, the $\phi^3$ theory is {\it nonperturbative} in $d = 4$. In particular, the singular behavior in the vicinity of the Lee-Yang point
\begin{equation}
  \label{eq:EOS_LY}
  M-M_\LY\sim (H-H_\LY)^{\sigma} ,
\end{equation}
is characterized by the exponent $\sigma\approx 0.26$ in $d = 4$ \cite{Gliozzi:2014jsa,Gracey:2015tta,An:2016lni,Zambelli:2016cbw}, which differs significantly from the mean-field result $\sigma = 1/2$.

We appear to be facing a paradox. On the one hand, mean-field theory should become valid as $d\to4$. On the other hand, this approximation fails to account for the correct exponent at the Lee-Yang point in the same limit. There is no contradiction, of course. The reason that mean-field theory becomes precise for $d\to4$ is that the importance of fluctuations diminishes as the fixed-point value of the coupling vanishes at $d = 4$. However, at any given value of $\epsilon$ (and $t$), the magnitude of the fluctuations themselves increase as we approach the Lee-Yang points, since the correlation length $\xi$ diverges at those points. In other words, the (squared) magnitude of fluctuations is proportional to the isothermal susceptibility $M'(H)$, which diverges as $H\to H_\LY$.

We are, therefore, led to seek a condition, similar to the Ginzburg criterion in the theory of superconductors \cite{Ginzburg:1960}, which determines how close the Lee-Yang edge singularity can be approached before mean-field theory breaks down. Even though the critical exponents, such as $\sigma$, cannot be determined reliably in the mean-field approximation, the domain of the validity of that approximation can be.

At the Lee-Yang point $H = H_\LY$ the mean-field potential Eq.\ \eqref{eq:V} takes the following form
\begin{equation}
  \label{eq:VLY}
  \left. V(\Phi)\right|_{H = H_\LY} = \frac{t^2}{12} + \frac{1}{\sqrt{3}}\, i t^{1/2} (\Phi-M_\LY)^3 + \frac{1}{4} (\Phi-M_\LY)^4 .
\end{equation}
It describes a {\it massless} $\phi^3$ theory with imaginary cubic coupling ($t > 0$). When $H\neq H_\LY$ a quadratic (mass) term appears. Expanding in $H-H_\LY$ we find
\begin{align}
  \label{eq:Vh-hLY}
  V(\Phi) & = \frac{t^2}{12} + (-3t)^{1/4} (H-H_\LY)^{1/2} (\Phi-M_\LY)^2 + \frac{1}{\sqrt{3}}\, i t^{1/2} (\Phi-M_\LY)^3 + \frac{1}{4} (\Phi-M_\LY)^4 + \ldots , \nonumber\\[-5pt] &
\end{align}
where we show only the leading-order contribution to each of the coefficients and the ellipsis denotes the subleading terms. From Eq.\ \eqref{eq:Vh-hLY} we can determine the correlation length, given by Eq.\ \eqref{eq:xi}, for small $H-H_\LY$. The result can be written in the following scaling form
\begin{equation}
  \label{eq:xi-h}
  \xi^{-2} = t \left[2 (-3)^{1/4} (w-w_\LY)^{1/2} + \ldots \,\right] ,
\end{equation}
where $w$ and $w_\LY$ are given by \eqref{eq:sv_wz} and \eqref{eq:zw_LY}, respectively. This analysis relies on the mean-field approximation and, therefore, assumes that fluctuations can be neglected.

The relative importance of fluctuations, is determined by the quartic coupling $\giv$, which is most evident in Eq.\ \eqref{eq:S2}, where $\giv$ controls the applicability of the saddle-point approximation to the path integral. In $4 - \epsilon$ dimensions, this coupling runs to the Wilson-Fisher fixed point in the IR, i.e., $\giv\to \givf = \ord(\epsilon)$ and therefore a mean-field (saddle-point) analysis is justified for sufficiently small $\epsilon$. How small $\epsilon$, or $\giv$, should be, however, depends on the value of the scaling-invariant variable $w$. For a generic value away from $w_\LY$ the condition is simply $\epsilon\ll 1$. However, as $w\to w_\LY$ the correlation length diverges, fluctuations are enhanced, and the condition on $\epsilon$ becomes more restrictive.

As $w\to w_\LY$, the relative importance of fluctuations is controlled by the most relevant coupling, the cubic coupling $\giii$, which can be read off as the coefficient of the $(\Phi-M)^3$ term in Eq.\ \eqref{eq:Vh-hLY}, i.e., $\giii\sim i (\giv t)^{1/2}$. Note that a factor $\sqrt{\giv}$ must be included in order to restore the canonical normalization of the field, $\Phi = \sqrt{\giv/6}\,\phi$. The mass dimension of the cubic coupling $\giii$ is $(6-d)/2$ and thus its relative importance is determined by the dimensionless combination $\giiit\equiv\giii\xi^{(6-d)/2}$ which, according to Eq.\ \eqref{eq:xi-h}, is given by 
\begin{equation}
  \label{eq:giiit-w}
  \giiit \sim \givt^{1/2}|w-w_\LY|^{-(6-d)/8} + \ldots ,
\end{equation}
where $\givt\equiv\giv t^{(d-4)/2} = \giv t^{-\epsilon/2}$. The mean-field analysis is applicable near the Lee-Yang edge singularity only as long as $\giiit\ll 1$.\footnote{The basic idea of the Ginzburg criterion is to compare the tree-level amplitude, or coupling, $\giii$ in our case, to the one-loop contribution. The latter stems from a triangle diagram, which is IR divergent when $\xi\to\infty$. By counting dimensions ($k^d$ from the loop integral and $k^6$ from the denominators) it is easy to see that the loop integral diverges as $\xi^{6-d}$. Thus, we need to compare $\giii$ to $(\giii)^3 \xi^{6-d}$, or equivalently $\giii^2 \xi^{6-d}$ to $1$.} For $0 < \epsilon\ll 1$, this yields the following requirement
\begin{equation}
  \label{eq:x-epsilon}
  |w - w_\LY|\gg \epsilon^2 ,
\end{equation}
where we replaced $\giv$ with its IR fixed-point value $\givf\sim\epsilon$. Eq.\ \eqref{eq:x-epsilon} is the Ginzburg criterion that determines the size of the critical region around the Lee-Yang point. Inside this region the mean-field approximation breaks down and the correct scaling near that point is given by the fixed point of the $\phi^3$ theory, which is nonperturbative in $d = 4$. One can also say that a typical condition for the mean-field approximation to apply, $\epsilon\ll1$, is not sufficient near the Lee-Yang points, where a stronger condition becomes necessary: Eq.\ \eqref{eq:x-epsilon}.

It is instructive to consider also the case $4 < d < 6$. The critical behavior simplifies as the scaling is now controlled by the Gaussian IR fixed point. However, we cannot simply set the coupling $\giv$ to zero since the action becomes singular in this limit (cf.\ Eq.\ \eqref{eq:S2}). In other words, for $d > 4$, the coupling $\giv$ is a dangerously irrelevant variable \cite{Amit:1982az}. In this case the equation of state depends on $\giv$ in addition to the variables $t$ and $H$. Repeating the arguments leading to Eq.\ \eqref{eq:giiit-w} we conclude that, for $4 < d < 6$, no matter how small the coupling $\giv$ is, the mean-field approximation will break down sufficiently close to the Lee-Yang point with the Ginzburg criterion given by
\begin{equation}
  \label{eq:w}
  |w-w_\LY|\gg (\givt)^{4/(6-d)} .
\end{equation}

Finally, for $d\ge 6$ the variable $w$ is not constrained by the Ginzburg criterion, and the condition $\givt\ll 1$ is sufficient for mean-field theory to apply for all $w$. This corresponds to the fact that $d = 6$ is the upper critical dimension of the $\phi^3$ theory, and the exponent $\sigma$ in Eq.\ \eqref{eq:EOS_LY} takes the mean-field value $1/2$ for $d\ge6$ in accordance with \cite{Fisher:1978pf}.

\section{Singularities in the $\epsilon$ expansion}
\label{sec:Singularities in the epsilon expansion}

\subsection{{Critical equation of state at $d = 4-\epsilon$}}
\label{sec:Critical equation of state at d = 4 - epsilon}

In this section we shall review the known results on the $\epsilon$
expansion of the equation of state relevant for our discussion.

Since the fixed-point value of the coupling $\giv$ is small near $d = 4$, the equation of state can be calculated perturbatively in $\epsilon = 4-d$ \cite{Brezin:1972fc,Avdeeva:1972,Wallace:1973,Wallace:1974,Nicoll:1985zz}. In terms of the rescaled variables $t$, $M$, and $H$, one finds to order $\epsilon^2$ \cite{Wallace:1973,Wallace:1974}\footnote{While the equation of state is known to order $\epsilon^3$ \cite{Wallace:1973,Wallace:1974,Nicoll:1985zz}, for our purposes it is sufficient to consider only contributions up to order $\epsilon^2$. We comment on some features specific to the $\epsilon^3$ result (in particular related to parametric representation of the equation of state) in Appendix \ref{sec:Parametric equation of state at order epsilon^3}.}
\begin{align}
  \label{eq:EOS_2nd}
  \frac{H}{M} = & ~ t + \frac{\gf}{8} r \left[ 1 + \ln(r) -\frac{\epsilon}{4} \ln^2(r) \right] - \frac{\gf^2}{64} r \left[ 4 + \pi^2 - 8 \lambda - \ln^2(r) \right] \nonumber\\ & +\: M^2 \left\{ 1 - \frac{3}{32} \gf^2 \left[ 6 + \frac{1}{2} \pi^2 - 4\lambda + 3 \ln(r) +\frac{1}{2} \ln^2(r) \right]\right\} ,
\end{align}
which reproduces the mean-field equation of state Eq.\ \eqref{eq:MFEOS} with Eq.\ \eqref{eq:tr0} when $\gf\to0$. Here, the parameter $\gf$ denotes the (Wilson-Fisher) fixed-point value of the conveniently normalized quartic coupling, i.e.,
\begin{equation}
  \label{eq:u}
  \gf = \givf\, \frac{S_d}{(2\pi)^d}\,\frac{\pi\epsilon}{\sin(\pi\epsilon/2)} = \frac{4}{3} \epsilon \left(1 + \frac{7}{54} \epsilon \right) + \ord(\epsilon^3) ,
\end{equation}
where $S_d$ is the area of the unit sphere in $d$ dimensions and $\lambda = (1/9)\left(3\Psi'(1/3) - 2 \pi^2\right)$ involves the first derivative of the digamma function $\Psi(z) = d/dz \ln \Gamma(z)$. The inverse isothermal susceptibility $r = (\partial H / \partial M)_t$ is a function of the variables $t$ and $M$, i.e., $r = r(t,M)$, and therefore Eq.\ \eqref{eq:EOS_2nd} constitutes a relation between $t$, $M$, and $H$.

As mentioned in Sec.\ \ref{sec:Introduction}, in the scaling region, Eq.\ \eqref{eq:EOS_2nd} can be written in a scaling form, in terms of scale-invariant combinations of two relevant variables. Among various choices it is more convenient for our work to write Eq.\ \eqref{eq:EOS_2nd} in terms of the scaling variables $w\sim H t^{-\beta\delta}$ and $z\sim M t^{-\beta}$, i.e., $w = F(z)$. Expressing the critical exponents $\beta$ and $\delta$ to order $\epsilon^2$, the ``gap'' exponent is given by
\begin{equation}
  \label{eq:Delta_epsilon}
  \beta\delta = \frac{3}{2} + \frac{1}{12} \epsilon^2 + \ord(\epsilon^3) ,
\end{equation}
and the series expansion in $\epsilon$ of the scaling function $F(z)$ reads
\begin{equation}
  \label{eq:w=F(z)_2nd}
  F(z) = F_0 (z) + F_1 (z) \epsilon + F_2 (z) \epsilon^2 + \ord(\epsilon^3) ,
\end{equation}
with
\begin{subequations}
  \label{eq:F012}
  \begin{align}
    \label{eq:F0} 
    F_0 (z) & = z + z^3 , \\
    F_1 (z) & = \frac{1}{6} \left[ -3z^3 + \left(z + 3z^3 \right)L(z)\right] , \\
    \label{eq:F2}
    F_2 (z) & = \frac{1}{648} \left[ -150 z^3 + 2(25 z - 6 z^3)L(z) + 9(z + 9 z^3)L^2(z)\right] ,
  \end{align}
\end{subequations}
and $L(z) = \ln\left[1+3z^2\right]$.\footnote{Note at the Lee-Yang point, the argument of the logarithmic function vanishes. An imaginary part, which is analyzed by Weinberg and Wu \cite{Weinberg:1987}, develops when the argument is negative and is associated with the cut terminating at the Lee-Yang edge singularity.} Note that the mean-field equation of state \eqref{eq:w=F(z)_MF} is recovered in the limit $\epsilon\to 0$. Here, the normalization of the scaling variables $w$ and $z$ in Eq.\ \eqref{eq:wz} is chosen in such a way that the two lowest order terms in the Taylor expansion
\begin{equation}
  \label{eq:PowerSeries}
  F(z) = z + z^3 + \sum_{n = 2}^{\infty} \mathcal{F}_{2n+1} z^{2n+1} ,
\end{equation}
are fixed and coefficients $\mathcal{F}_{2n +1} = \ord(\epsilon)$, for all $n\geq 2$.

Since the singularities of the equation of state are associated with a diverging correlation length, the equation of state must be analytic away from the Ising critical point located at $t = M = H = 0$. Thus, if any one of these parameters is set to a nonzero value, the relation between the other two must be analytic. This translates into the following two properties of $F(z)$ often referred to as Griffiths' analyticity \cite{Griffiths:1967}. First, for fixed $t > 0$, we find that $F(z)\sim H$ is an analytic function of $z\sim M$ in the vicinity of $z = 0$, which should also be odd under reflection $H\to -H$ and $M\to -M$. This is easily seen in the explicit expressions for $F(z)$ in Eqs.\ \eqref{eq:F012}. Second, for fixed $M > 0$ we find that the function $z^{-\delta}F(z)\sim H$ must be an analytic function of the variable $z^{-1/\beta}\sim t$ in the vicinity of $t=0$ ($z=\infty$). The behavior of $F(z)$ at large $z$ is not manifest in Eqs.\ \eqref{eq:F012} since the $\epsilon$ expansion of this function does not converge uniformly, due to the presence of large logarithms. 

In this case, it is better to introduce the scaling variables $x \sim t M^{-1/\beta}$ and $y \sim H M^{-\delta}$ (as in Eq.\ \eqref{eq:xy}), and express the equation of state Eq.\ \eqref{eq:EOS_2nd} as the Widom scaling function $y=f(x)$ \cite{Guida:1996ep}, whose $\epsilon$ expansion is convergent when $x\to0$ (corresponding to $z\sim x^{-\beta}\to\infty$). However, in this representation the analyticity at large $x$ (corresponding to small $z$) is obscured, again due to lack of convergence of the $\epsilon$ expansion.

Thus it would be useful to have a representation of the equation of state where the analyticity is manifest in both regimes, i.e., a representation for which the $\epsilon$ expansion converges uniformly. The so-called parametric representations \cite{Schofield:1969,Josephson:1969}, reviewed below, are designed to fulfill this requirement.

\subsection{Parametric equation of state}
\label{sec:Parametric equation of state}

As we discussed in the previous section, the problem with representations using the pairs of scaling variables such as $w$ and $z$, or $y$ and $x$, is that the two points $z = 0$ and $x\sim z^{-1/\beta} \to 0$, where each of them is analytic, correspond to infinitely separated points $z = 0$ and $z= \infty$ （and similarly for $x$）. This problem can be addressed by introducing a new scaling variable, $\theta$, by means of a nonlinear variable transformation $(t, M) \to (R, \theta)$:
\begin{align}
  \label{eq:ParametricModel1}
  t & = R \hspace{1pt} k(\theta) , \\
  \label{eq:ParametricModel2}
  M & = R^{\beta} m(\theta) ,
\end{align}
with {\it analytic} functions $k(\theta)$ and $m(\theta)$, chosen such that the two points $x\sim t M^{-1/\beta} = 0$ and $z\sim M t^{-\beta} = 0$ are placed at positions $\theta = 1$ and $\theta = 0$, respectively. The simplest choice satisfying these conditions is
\begin{equation}
  k(\theta) = 1-\theta^2 \quad\textrm{and}\quad m(\theta) = \bar{m} \theta ,
\end{equation}
also known as the linear parametric model (LPM) \cite{Schofield:1969,Schofield:1969zz}. Here, $\bar{m}$ is a normalization constant, which can be chosen to bring the equation of state into canonical form (e.g., see Eq.\ \eqref{eq:PowerSeries}).

In the parametric representation, the equation of state becomes a relationship between $H$ and the parameters $R$ and $\theta$, i.e.,
\begin{equation}
  \label{eq:ParametricModel3}
  H = R^{\beta\delta} h(\theta) ,
\end{equation}
where $h(\theta)$ is an odd function of $\theta$ (since $\theta\sim M$ is an odd variable under reflection $M\to -M$, $H\to -H$). 

While $R$ scales as the reduced temperature $t$, the variable $\theta$
is invariant under rescaling in Eq.\ \eqref{eq:scaling}. Therefore, the scaling variables $w$ and $z$ can be expressed in terms of $\theta$ alone, i.e.,
\begin{equation}
  \label{eq:wz-theta}
  z \sim M t^{-\beta} \sim \theta (1-\theta^2)^{-\beta} \quad {\rm and} \quad
  w \sim H t^{-\beta\delta} \sim h(\theta) (1-\theta^2)^{-\beta\delta} .
\end{equation}
Inserting these expressions into the equation of state $w = F(z)$ one can determine the function $h(\theta)$ (as well as the normalization constant $\bar{m}$) order by order in the $\epsilon$ expansion \cite{Guida:1996ep}.\footnote{Similarly, one could also use the equation of state in the form $y = f(x)$ with the scaling variables $x\sim \theta^{-1/\beta} (1-\theta^2)$ and $y\sim \theta^{-\delta} h(\theta)$ to determine $h(\theta)$.}

In this section, we shall carefully examine the parametric representation obtained by matching equation of state to order $\epsilon^2$. Our goal is to determine the location of singularities and their uncertainty due to higher orders of $\epsilon$ expansion. To focus on relevant features we present the results in the minimal form necessary for the argument, and collect explicit expressions needed for the derivation in Appendix \ref{sec:Details of the parametric equation of state at order epsilon^2} and \ref{sec:Parametric equation of state at order epsilon^3}.

It is known that to order $\epsilon^2$ the function $h(\theta$) is given by a cubic
polynomial \cite{Brezin:1972fc,Wallace:1973,Wallace:1974}
\begin{equation}
  h(\theta) = \bar{h}(\theta + h_3 \theta^3 ) ,
  \label{eq:LPM2}
\end{equation}
where $\bar h$ is a normalization parameter. As we shall see, the number of singularities is determined by the order of this polynomial while their positions are related to the coefficient $h_3$ which can be determined by matching to equation of state \eqref{eq:PowerSeries}. For $\epsilon=0$ (mean-field equation of state), $h_3=-2/3$.

In order to study the dependence of our results on $\epsilon$ we shall expand $h_3$ in $\epsilon$. To this end we adopt the historical notation of Refs.\ \cite{Schofield:1969,Schofield:1969zz,Wallace:1973,Wallace:1974} and express $h_3$ in terms of parameter $b$ defined by $h(\theta = b) = 0$, i.e., the closest zero to $\theta = 0$. Obviously,
\begin{equation}
  \label{eq:h_3_2nd}
  h_3 = -\frac{1}{b^2}.
\end{equation}
The coefficients of the $\epsilon$ expansion of $b^2$ 
\begin{equation}
  b^2 = \frac{3}{2} + b_1\epsilon + b_2\epsilon^2 + \ord(\epsilon^2) .
\end{equation}
cannot be determined by matching at order $\epsilon^2$ (or $\epsilon^3$ for that matter, cf.\ Appendix \ref{sec:Parametric equation of state at order epsilon^3}). It is a common choice \cite{Wallace:1973,Wallace:1974} to set $b_1 = 0$, but it is not necessary and we shall allow this parameter to have an arbitrary real value. It will be helpful for understanding the $\epsilon$ dependence of our results.

We shall now study the singularities that arise in the linear parametric representation in order to infer the analytic properties of the scaling equation of state. Specifically, we examine the equation of state to order $\epsilon^2$ in the form $w = F(z)$, represented parametrically using Eqs.\ \eqref{eq:wz-theta}. This allows us to directly access the singularities in the complex $w$ plane by examining the rescaled inverse isothermal susceptibility, given by
\begin{equation}
  \label{eq:rwz}
  r t^{-\gamma} \sim F'(z) = \frac{w'(\theta)}{ z'(\theta) } ,
\end{equation}
whose zeros correspond to the branching points of the multivalued function $z(w)$.

In terms of the linear parametric representation, Eqs.\ \eqref{eq:ParametricModel1} -- \eqref{eq:ParametricModel3} and Eq.\ \eqref{eq:LPM2}, the scaling variables $z$ and $w$ are given by
\begin{equation}
  \label{eq:wz-theta-LPM}
  z = \frac{\bar{z}\hspace{1pt} \theta}{(1-\theta^2)^{\beta}} \quad {\rm and} \quad
  w = \frac{\bar{w}\hspace{1pt} (\theta + h_3 \theta^3)}{(1-\theta^2)^{\beta\delta}} ,
\end{equation}
where the normalization parameters $\bar{z}$ and $\bar{w}$, determined by matching the parametric model to the canonical equation of state Eq.\ \eqref{eq:PowerSeries} to order $\epsilon^2$, are needed below to find the position of singularities to that order and are given by Eqs.\ \eqref{eq:zwbar_2nd_full}. Substituting into Eq.\ \eqref{eq:rwz} we arrive at the following expression for the inverse susceptibility
\begin{equation}
  \label{eq:F'z-LPM}
  F'(\theta) = \frac{\bar{w}}{\bar{z}} (1-\theta^2)^{-\gamma} \frac{1 + \left(2 \beta\delta +3 h_3-1\right) \theta^2 + (2 \beta\delta -3) h_3 \theta^4 }{1- (1-2\beta)\hspace{1pt} \theta^2} ,
\end{equation}
where the scaling exponents $\beta$, $\gamma$, and $\delta$, as well as the parameters $h_3$, $\bar w$ and $\bar z$ should be expanded to order $\epsilon^2$.

If we set $\epsilon = 0$ and use the mean-field critical exponents, $\beta = 1/2$, $\gamma = 1$, and $\beta\delta = 3/2$, we observe that the only zeros of $F'(\theta) = (1-\theta^2)^{-1}$ lie at complex infinity (in the $\theta$ plane). Of course, this is consistent with the mean-field result, which is easily confirmed by examining the limit $|\theta|\to \infty$ in Eqs.\ \eqref{eq:wz-theta-LPM}, i.e., $\lim_{|\theta|\to\infty} w(\theta) = \pm 2 i/ (3\sqrt{3})$, and comparing with \mbox{Eq.\ \eqref{eq:zw_LY}}.

At nonzero $\epsilon$, however, the structure of the singularities of Eq.\ \eqref{eq:F'z-LPM} becomes more complicated. Now, the polynomial in the numerator has four zeros. There are also two zeros in the denominator, giving rise to two poles. In addition, there are two branch-point singularities at $\theta = \pm 1$. Since $w(\theta = 1) = z(\theta = 1) = \infty$ the latter can be seen to correspond to the behavior $F(z)\sim z^\delta$ (and, therefore, $F'(z)\sim z^{\gamma/\beta}$) at large $z$, required by Griffiths' analyticity. The four zeros and two poles on the other hand, occur at finite, albeit large, values of $\theta^2 = \ord(\epsilon^{-1})$. We shall now focus on these finite $w$ singularities.

Since $F'(\theta)$ is an even function of $\theta$ it is convenient to consider its singularities as a function of $\theta^2$. The numerator of Eq.\ \eqref{eq:F'z-LPM} vanishes at two distinct values $\theta_n^2$, which we label by indices $n = 1,2$. {These solutions can be expanded in powers of $\epsilon$ where the leading contribution appears at order $\epsilon^{-1}$, i.e.,
\begin{equation}
  \label{eq:SeriesAnsatz1}
  \theta_n^{2} = \frac{c_n}{\epsilon}\left[ 1 + \ord(\epsilon) \right] . 
\end{equation}
Substituting $\theta_n$ into Eq.\ \eqref{eq:wz-theta-LPM}, $w_n\equiv w(\theta_n)$, and expanding in $\epsilon$ we get
\begin{eqnarray}
  \label{eq:w_n_2nd}
  \hspace{0pt} w_n =\pm \frac{2 i \left(-\hat{c}_n\right)^{\frac{3}{2} - \beta\delta}}{3\sqrt{3}} \left\{ 1 + \left[\omega^{(2)}(c_n, b_1)+ \frac{1}{12}\ln\epsilon \right]\epsilon^2 + \ord(\epsilon^3) \right\} ,
\end{eqnarray}
where $\hat{c}_n\equiv c_n/|c_n|$. Remarkably, only the leading-order coefficient of $\theta_n^2$, $c_n$, appears in this expression.\footnote{This happens because the leading corrections to the mean-field value of $w$ are $\epsilon\theta^{-2}$ and $\theta^{-4}$, while $\theta_n^{-2}\sim\epsilon$.} The coefficient $c_n$ is a function of $b_1$ and for the two solutions $\theta_n^2$, $n = 1,2$, we obtain
\begin{equation}
  \label{eq:SolutionsKappa1}
  c_n = 3 \left( 2 b_1 + (-1)^{n} \sqrt{1 + 4 b_1^2} \right), \quad n = 1,2 ,
\end{equation}
with $c_1 < 0$ and $c_2 > 0$ for all real values of $b_1$. Note that the absolute value of $w_n$ is determined by $\omega^{(2)}(c_n(b_1), b_1)$ -- a function of $b_1$ (see Eq.\ \eqref{eq:w_n_2nd_full}) while the dependence on $n$ appears only via $c_n$ in Eq.\ \eqref{eq:SolutionsKappa1}. Nontrivially, there are no $\mathcal{O}(\epsilon)$ terms in Eq.\ \eqref{eq:w_n_2nd} (they cancel) and there is no dependence on $b_2$ to this order.

\begin{figure}[!t]
  \centering
  \includegraphics[width = 0.3\textwidth]{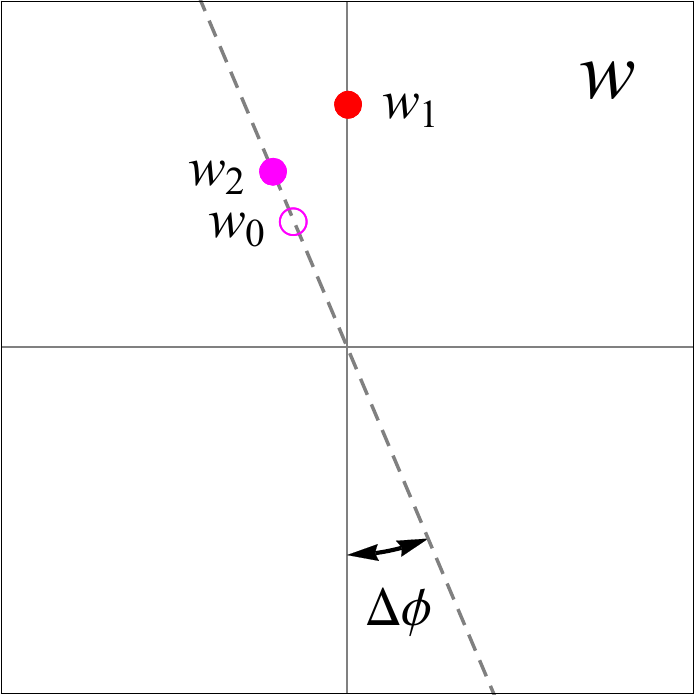}
  \caption{\label{fig:5}We show the position of the two zeros $w_1$ and $w_2$ (solid points) and single pole $w_0$ (open circle) of the parametrically represented inverse isothermal susceptibility $F'(z)$ at order $\ord(\epsilon^2)$ in the complex $w$ plane ($b_1\neq 0$). Note, only the singularities in the upper half of the complex $w$ plane are shown.}
\end{figure}

Inserting the coefficient $c_1$ into \eqref{eq:w_n_2nd}, we find
\begin{equation}
  w_1 = \pm \frac{2 i}{3\sqrt{3}} \left\{ 1 + \left[\omega^{(2)}(c_1, b_1)+ \frac{1}{12}\ln\epsilon \right]\epsilon^2 + \ord(\epsilon^3) \right\} ,
\end{equation}
which corresponds to the pair of Lee-Yang edge singularities (cf.\ Eq.\ \eqref{eq:zw_LY}). As in the mean-field case they are located on the imaginary axis in accordance with Lee-Yang theorem. However, comparing with the mean-field result, we observe that its absolute value receives corrections of order $\epsilon^2$, which depend on the parameter $b_1$. Since $b_1$ cannot be determined at this order of the $\epsilon$ expansion, practically, the position of the singularity also cannot be established to precision of order $\epsilon^2$. This agrees with our earlier observation in Sec.\ \ref{sec:Lee-Yang edge singularities and Ginzburg criterion} that the nonperturbative domain around the Lee-Yang edge singularities has size $\ord(\epsilon^2$), according to Ginzburg criterion Eq.\ \eqref{eq:x-epsilon}.

The second pair of singularities, $w_2$, is located off the imaginary axis:
\begin{equation}
  \label{eq:w_1_2nd}
  w_2 = \pm \frac{2 i}{3\sqrt{3}} (-1)^{\frac{3}{2} - \beta\delta} \left\{ 1 + \left[\omega^{(2)}(c_2, b_1)+ \frac{1}{12}\ln\epsilon \right]\epsilon^2 + \ord(\epsilon^3) \right\} .
\end{equation}
One can easily see that they lie precisely where we expect the Langer cut (see Fig.\ \ref{fig:3}). But what is their significance? Before answering this question,  let us first consider the poles of $F'(z)$, which can be obtained by solving
\begin{equation}
  \label{eq:Zeros2}
  1 - \left(1 - 2\beta \right) \theta^2 = 0 .
\end{equation}
The solution $\theta^2$ to this equation, which we label by the index $n = 0$, can be also expanded in powers of $\epsilon$. The corresponding leading coefficient $c_0$ (cf.\ Eq.\ \eqref{eq:SeriesAnsatz1}) is given by
\begin{equation}
  \label{eq:SolutionsKappa1b}
  c_0 = 3 ,
\end{equation}
and according to Eq.\ \eqref{eq:w_n_2nd}, we find 
\begin{eqnarray}
  \label{eq:w_2_2nd}
  w_0 = \pm \frac{2 i}{3\sqrt{3}} (-1)^{\frac{3}{2} - \beta\delta} \left\{ 1 + \left[\omega^{(2)}(c_0, b_1)+ \frac{1}{12}\ln\epsilon \right]\epsilon^2 + \ord(\epsilon^3) \right\} .
\end{eqnarray}}

The position of the singularities $w_0$, $w_1$, and $w_2$, is shown schematically in Fig.\ \ref{fig:5} in the upper half of the complex $w$ plane and for a generic value of $b_1$, according to Eq.\ \eqref{eq:w_n_2nd_full}. We observe that $w_0$ and $w_2$ lie on the same ray, which corresponds to the Langer cut of the {\it exact} equation of state. The distance between these points is given by
\begin{equation}
  w_2 - w_0 = \ord(\epsilon^2) ,
\end{equation}
and depends on the value of $b_1$. For the common and particular choice $b_1 = 0$, when $c_2=c_0$, the two points coincide and the zero and the pole cancel each other to order $\epsilon^2$.

It is also important to note that both $w_0$ and $w_2$ (on the Langer cut) are within distance $\ord(\epsilon^2)$ from the Lee-Yang edge singularity, since $\beta\delta-3/2 = \ord(\epsilon^2)$. Therefore, according to the Ginzburg criterion in Eq.\ \eqref{eq:x-epsilon}, these singularities and their position are nonperturbative. This is in agreement with the fact that we cannot determine the parameter $b_1$ within the $\epsilon$ expansion to establish their position. Furthermore, as we show in Appendix \ref{sec:Parametric equation of state at order epsilon^3}, extending the linear parametric model to next order, $\epsilon^3$, leads to terms in $w_n$ that contribute at order $\epsilon^2$ (in {\it addition} to the expected $\epsilon^3$ contribution). Thus, the procedure based on matching to increasing orders of the $\epsilon$ expansion does not converge in the usual sense. In spite of this, it is still tempting to speculate that the sequence of (alternating) zeros and poles line up along the ray at angle $\Delta\phi$ relative to the imaginary axis and will eventually coalesce into the Langer cut -- a purely nonperturbative feature, which cannot be reproduced at any finite order of $\epsilon$ expansion. In fact, such a scenario is common in rational-function (Pad\'e) approximations of functions with branch cuts.\footnote{The experience with Pad\'e approximations suggests a guiding principle for constructing improved parametric representations: The choice of (polynomial) functions $h(\theta)$ and $m(\theta)$ should be such that the rank of polynomials in the numerator and the denominator in Eq.\ \eqref{eq:F'z-LPM} increase at the same rate.}

Summarizing, we see that the Ginzburg criterion \eqref{eq:x-epsilon} sets the limit on the information that can be gained about the Lee-Yang edge singularities. The precision that we can reach, $\epsilon^2$, is not sufficient to study the region between the Lee-Yang edge singularity and the Langer cut, which is necessary to test the Fonseca-Zamolodchikov conjecture. Nevertheless, the results we find are nontrivially consistent with the conjecture.

\section{Singularities in the $\O(N)$-symmetric $\phi^4$ theory}
\label{sec:Singularities in the O(N)-symmetric phi^4 theory}

An alternative point of view on the question of extended analyticity and the nature of singularities in the complex $H$ plane can be obtained by studying the generalization of the $\phi^4$ theory to the $N$-component theory with $\O(N)$ global symmetry. This generalization is a well-known tool to study nonperturbative aspects of the theory. The finite-$N$ cases describe the critical behavior of, e.g, the Heisenberg model ($N = 3$), the XY-model ($N = 2$), and, of course, the Ising model ($N = 1$). On the other hand, in the $N\to \infty$ limit the $\O(N)$ model describes the critical behavior of the exactly solvable spherical model \cite{Berlin:1952zz,Stanley:1968gx}.

Similar to Eq.\ \eqref{eq:S}, the $\O(N)$ theory is defined by the Euclidean action (or Hamiltonian divided by temperature)
\begin{equation}
  \label{eq:SOn}
  \mathcal{S} = \int d^dx \left[\frac{1}{2} (\partial_\mu\boldsymbol{\phi})^2 + \frac{r_0}{2} \boldsymbol{\phi}^2 + \frac{\giv}{4!}(\boldsymbol{\phi}^2)^2 - \boldsymbol {h_0\cdot \phi} \right] .
\end{equation}
Here, $\boldsymbol{\phi}$ is a (real) $N$-component vector field and the external magnetic field $\boldsymbol{h_0}$ has the same dimensionality. In the presence of a nonvanishing $\boldsymbol{h_0}$ the expectation value of $\boldsymbol{\phi}$, $\langle\boldsymbol\phi\rangle$, is directed along the former. Due to the $\O(N)$ invariance of the theory we may choose an axis along the vector $\bm h_0$ and define the equation of state as a relationship between the projections $\langle\phi\rangle$, $h_0$ of $\langle\boldsymbol\phi\rangle$ and $\bm h_0$ onto that direction,
similar to the $N = 1$ case in Sec.\ \ref{sec:Critical equation of state and the mean-field approximation}.

We are interested in analytic properties of the universal equation of state, which describes the critical behavior of the $\phi^4$ theory associated with the spontaneous breaking of the $\O(N)$ symmetry. However, since there can be no spontaneous symmetry breaking of continuous symmetries in $d\leq 2$ \cite{Mermin:1966fe,Coleman:1973ci}, we shall limit our analysis to dimensions $d > 2$.

When $h_0 = 0$ the critical point is reached by tuning $r_0$ to its critical value, i.e., $t = r_0 - r_\c\to 0$. In this limit, and for $d < 4$ the quartic coupling $\giv$ runs into the $\O(N)$ Wilson-Fisher fixed point in the infrared, i.e., $\giv\to \givf$ \cite{Brezin:1972fb,Brezin:1972se} and therefore the critical equation of state becomes independent of the bare coupling $\giv$ as well as the ultraviolet cutoff. A systematic expansion in powers of $1/N$, yields a fixed-point value with $\givf\sim \ord(1/N)$ \cite{Brezin:1972se,Abe:1977}. But this does not necessarily mean that the equation of state of the $\O(N)$ model reduces to the mean-field result in the limit $N\to \infty$. Indeed, the tree-level action for the longitudinal field $\phi$ receives a nontrivial contribution from integrating out the $N-1$ transverse-field degrees of freedom (which, for $t < 0$, correspond to the massless Goldstone modes associated with the spontaneous breaking of the $\O(N)$ symmetry) \cite{Brezin:1972se}. Both the tree-level action, proportional to $1/\giv\sim \ord(N)$ (as in Eq.\ \eqref{eq:S2}), as well as the one-loop contribution of the $N-1$ transverse-field modes are of order $N$. We may therefore apply the saddle-point approximation in the large-$N$ limit.

We shall first consider the infinite-$N$ case, or the spherical model, and then briefly comment on $1/N$ corrections below. As in Sec.\ \ref{sec:Critical equation of state and the mean-field approximation} we introduce the rescaled field variables $M = \sqrt{\giv/6}\,\langle\phi\rangle$ and $H = \sqrt{\giv/6}\,h_0$ and employ the scaling variables $w = H t^{-\beta\delta}$ and $z = M t^{-\beta}$ etc.

In the $N\to \infty$ limit the critical exponents are known \cite{Brezin:1972se}
\begin{equation}
  \label{eq:exponents-N}
  \beta = \frac{1}{2} , \quad \gamma = \frac{2}{d-2} , \quad{\rm and}\quad \delta = \frac{d+2}{d-2} , \quad {\rm for} \quad 2 < d < 4 ,
\end{equation}
and take their mean-field values for $d\geq 4$. The scaling equation of state $w = F(z)$ is determined in terms of the scaling function
\begin{equation}
  \label{eq:w_largeN}
  F(z) = z (1+z^2)^{\gamma} .
\end{equation}
In $d\geq 4$ dimensions, where the critical exponent $\gamma = 1$, this agrees with the mean-field equation of state Eq.\ \eqref{eq:w=F(z)_MF}, as should be expected.

The branching points of the inverse function $z(w)$ correspond to solutions of $F'(z) = 0$. We find two (pairs of) such solutions
\begin{equation}
  \label{eq:2}
  z^2 = -1 \quad\mbox{and}\quad z^2 = -\frac{1}{1+2\gamma} , 
\end{equation}
which map onto 
\begin{equation}
  \label{eq:branching-largeN}
  \qquad w = 0 \quad {\rm and}\quad w = \pm i (2\gamma)^{\gamma} (1+2\gamma)^{-\beta\delta} ,
\end{equation}
in the complex $w$ plane.

The $w\neq0$ solutions lie on the imaginary $w$ axis. In fact, for $d = 4$ they are {\it identical} to the Lee-Yang edge singularities in Eq.\ \eqref{eq:LY_MF}. Thus, for $t > 0$, we can identify these solutions with the pair of Lee-Yang edge singularities at imaginary $H$. For $t < 0$, they lie on the real $H$ axis for $d\geq 4$, while they are shifted off the real $H$ axis by an angle\footnote{Note, the angular displacement $\Delta\phi$ is of order $\epsilon$ and {\it not} $\epsilon^2$ as in the Ising-like ($N = 1$) case (cf.\ Sec.\ \ref{sec:Langer cut and Fonseca-Zamolodchikov conjecture}).}
\begin{equation}
  \label{eq:1}
  \Delta\phi = \pi \frac{4-d}{d-2} > 0 , \quad {\rm for}\quad 2 < d < 4 ,
\end{equation}
as expected, cf.\ Eq.\ \eqref{eq:beta-delta-32}.

But what is the meaning of the solution at $w = 0$ (i.e., $H = 0$) in Eq.\ \eqref{eq:branching-largeN}? Since $z^2 = -1$ and $M\sim z t^{1/2}$, this singularity corresponds to real $M$ only for $t < 0$. It is located at the origin ($H = 0$) of the low-temperature sheet and is associated with a branch cut along the negative real $H$ axis.

\begin{figure}[!t] 
  \begin{picture}(0,120)
    \put(0,0){\includegraphics[width = 0.455\textwidth]{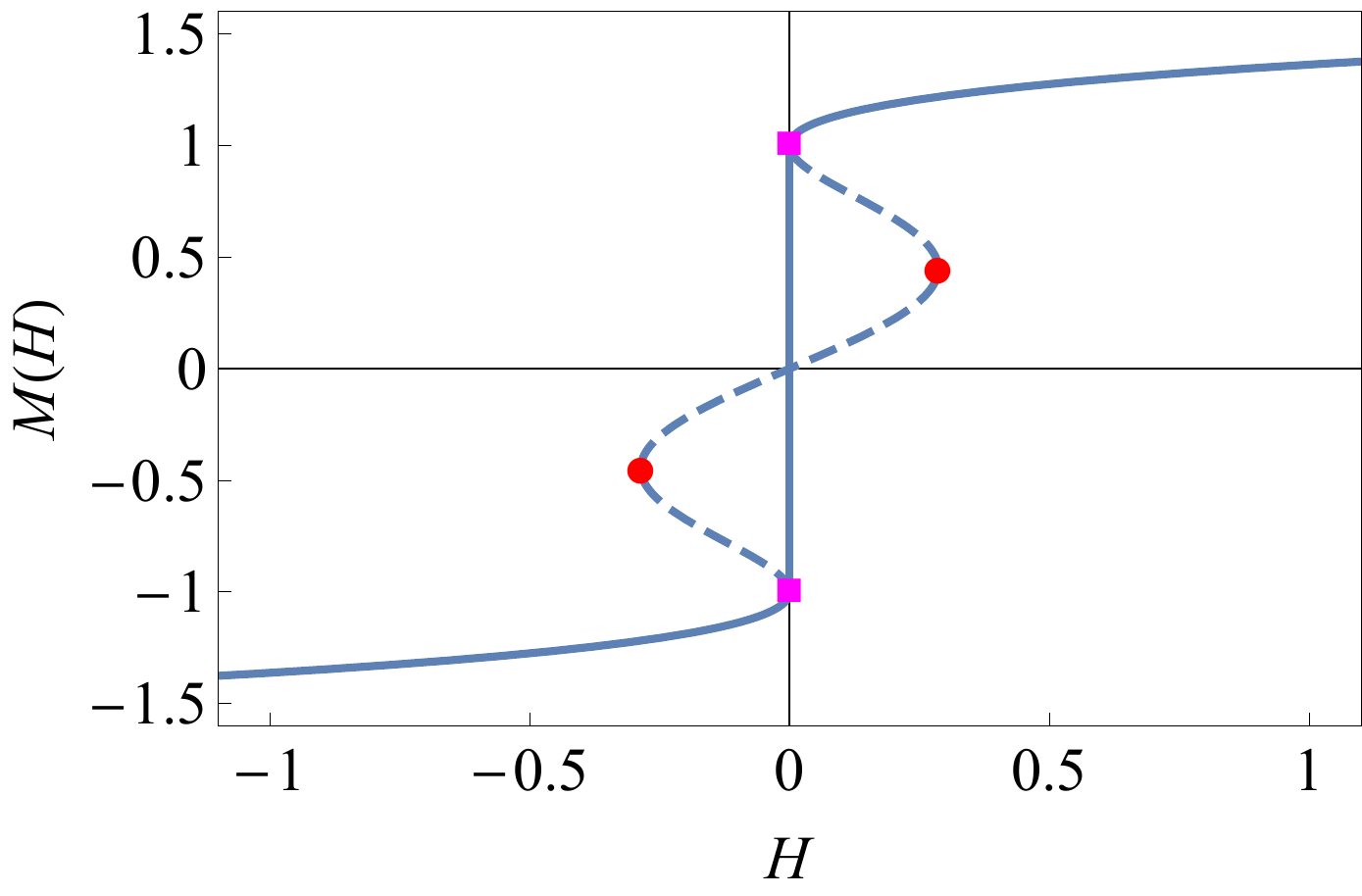}}
    \put(235,0){\includegraphics[width = 0.45\textwidth]{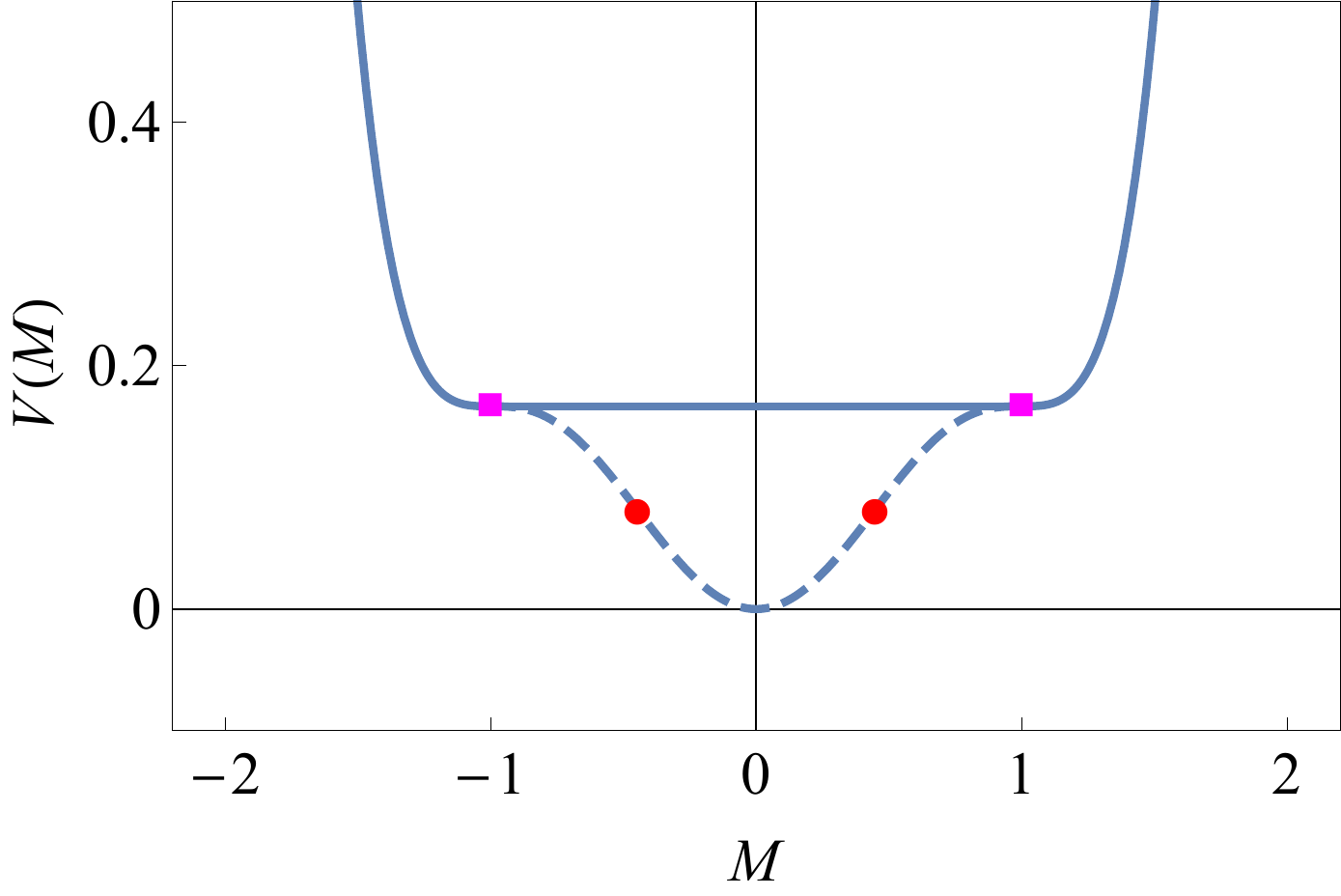}}
    \put(40,124){(a)}
    \put(268,124){(b)}
  \end{picture}
  \caption{\label{fig:6}(a) Equation of state $M(H)$ of the three-dimensional $\O(N)$ model in the $N\to \infty$ limit and (b) the corresponding effective potential $V(M)$ at $H = 0$ in the low-temperature phase ($T < T_\c$). The dashed curve illustrates the analytic continuation of the stable branch (solid curve). In addition to the spinodal singularities at nonvanishing $H$, the presence of massless Goldstone modes induces singularities on the coexistence line ($T < T_\c$ and $H\to 0$).}
\end{figure}

To understand the significance of this singularity and the associated branch cut, we illustrate the equation of state $M(H)$ in Fig.\ \ref{fig:6} for $t < 0$ and $d = 3$. Unlike the $N = 1$ case there is no metastable regime (compare with Fig.\ \ref{fig:1}). This can be understood as a consequence of the fact that, when $H$ changes sign (relative to $M$), the effective potential as a function of $\bm\phi$ develops directions with negative curvature (i.e., the Goldstone bosons become tachyonic). This means that the false vacuum is classically unstable and, since there is no tunneling involved in the decay of the false vacuum, there is also no exponential suppression of the imaginary part, unlike the $N = 1$ case. That is, instead of the essential (and very weak) singularity (cf.\ Eq.\ \eqref{eq:LangerSingularity}) the equation of state with $N > 1$ has a {\it power-law} singularity \cite{Brezin:1972fb,Brezin:1972se,Wallace:1975,Gunther:1980} which comes from the IR-divergent contributions of the Goldstone bosons \cite{Wallace:1975,Lawrie:1981}. Similar to the Langer cut in the Ising case, the $N > 1$ equation of state for $2 < d < 4$ has a ``Goldstone cut'' branching off from the origin and going along the real $H$ axis on the unstable branch ($H < 0$ in our convention), with discontinuity given by\footnote{The coefficient of this singularity vanishes at $N = 1$ \cite{Wallace:1975}.}
\begin{equation}
  \begin{split}
    \label{eq:GoldstoneSingularity}
    & \im M\sim H^{(d-2)/2}, \quad {\rm for} ~ H\to 0 , \quad t < 0 .
  \end{split}
\end{equation}

Furthermore, from Fig.\ \ref{fig:6}, we see that, for $T < T_\c$, the Lee-Yang edge singularities must lie on another (unphysical) branch of the equation of state $M(H)$. These singularities can be reached in the complex $H$ plane by going under the Goldstone cut onto an ancillary Riemann sheet. In fact, the situation is very similar to the conjectured scenario shown in Fig.\ \ref{fig:4}, where we observed a similar analytic structure of the equation of state.\footnote{Interestingly, the analytic structure of the scaling equation of state of the three-dimensional spherical model is also remarkably similar to that of the planar Ising model coupled to two-dimensional quantum gravity \cite{Kazakov:1986hu,Boulatov:1986sb,Bourgine:2011qp}.} The low-temperature singularities located off the Goldstone cut for $d < 4$ are very similar to the spinodal points. In fact, they become the spinodal points at $d = 4$ when the equation of state \eqref{eq:w_largeN} takes the mean-field form \eqref{eq:w=F(z)_MF}.

Since there are no singularities in the equation of state Eq.\ \eqref{eq:w_largeN} apart from the ones given by Eqs.\ \eqref{eq:branching-largeN}, we conclude that the Fonseca-Zamolodchikov scenario is realized in the $\O(N)$ model in the $N\to \infty$ limit.

To complete our analysis, we finally comment on the $1/N$ corrections to the scaling function \eqref{eq:w_largeN}. Since the leading-order contribution to $\Delta\phi$ in Eq.\ \eqref{eq:1} is already $\ord(1/N^0)$, we observe that $1/N$ corrections cannot change the conclusion that there are no singularities at real (nonzero) $H$, provided that the only effect of these corrections is to shift the position of the singularities already present in the $N\to\infty$ limit.

The $1/N$ corrections can be expressed in terms of momentum integrals whose explicit form is not particularly illuminating (see Refs.\ \cite{Brezin:1972se,Abe:1977} for details). For simplicity we shall consider only $d = 3$, which is also the case that is most relevant for applications. In three dimensions, we find that the aforementioned momentum integrals yield only two branch points at $z^2 = -1$ and $z^2 = -1/5$, which coincide with the same singularities already found in the $N\to \infty$ limit, cf.\ Eq.\ \eqref{eq:2}, while the position of the corresponding points in the complex $w$ plane is shifted by an amount of order $1/N$.\footnote{At order $1/N$ the value of $\Delta\phi$, which controls the position of the spinodal points in the complex $w$ (or $H$ plane) is given by $\Delta\phi = \pi\left(\beta\delta -\frac{3}{2}\right) = \pi - {28}/({\pi N}) + \ord(1/N^2)$ in $d = 3$ dimensions, where we have used Eq.\ \eqref{eq:beta-delta-32} and critical exponents $\beta$ and $\delta$ from Ref.\ \cite{Abe:1972,Brezin:1972se}.} This is consistent with our expectation that the $1/N$ corrections only modify the position of the singularities (as determined in the $N\to\infty$ limit) and suggests that no new singularities appear at finite $N$.\footnote{Note that the position of the singularities in the complex $w$ plane can be calculated reliably to order $1/N$, even though higher-order corrections become nonperturbative. Indeed, the Ginzburg criterion (see Eq.\ \eqref{eq:giiit-w} or \eqref{eq:w}) with $\giv = \ord(1/N)$ imposes a constraint on the applicability of the saddle-point approximation around the Lee-Yang point, which reads $|w-w_\LY|\gg N^{-4/(6-d)}$ (in agreement with Ref.\ \cite{Bander:1984}). This demonstrates that the nonperturbative region is {\it smaller} than $1/N$ for $d > 2$.}

\section{Conclusions}
\label{sec:Conclusions}

In this work we studied the relationship between singularities of the universal scaling equation of state of the $\phi^4$ theory above and below the critical temperature. Above the critical temperature Lee-Yang edge singularities, by the Lee-Yang theorem, lie on the imaginary magnetic field axis and limit the domain of analyticity around the origin $H = 0$. On the other hand, below the critical temperature, there are singularities associated with the point where the metastable state becomes locally unstable and its decay occurs via spinodal decomposition. 

In the mean-field approximation to the equation of state, $H = t M + M^3$, these spinodal points are related to the Lee-Yang edge singularities. In terms of the scaling variable $w = H t^{-\beta\delta}$, they are essentially the same singularities. These singularities occur at imaginary $w$ and, for $t > 0$, they correspond to imaginary $H$, i.e., the Lee-Yang points. For $t < 0$, however, they correspond to {\it real} $H$ on the metastable branch (since in the mean-field approximation: $\beta\delta = 3/2$ and $i(-1)^{3/2} = -1$).

Since $\beta\delta\neq 3/2$ for $d < 4$, one naturally has to ask the question if the spinodal singularities on the real $H$ axis exist at all. The analyticity of equation of state as a function of $w$ would require the low-temperature manifestation of the Lee-Yang points to be points off the real axis by a phase $\Delta\phi = \pi(\beta\delta-3/2)$. Fonseca and Zamolodchikov put forward a conjecture that these are the closest singularities to the real $H$ axis. Our aim here was to test this conjecture in the small-$\epsilon$ and large-$N$ regimes.

We have used a uniform approximation to the equation of state based on parametric representations, which are especially convenient to study the equation of state in the whole complex plane of $w$ using the $\epsilon$ expansion. However, the vicinity of the Lee-Yang singularity is special in that the $\epsilon$ expansion must break down. In fact, there is an apparent paradox, identified first by Fisher \cite{Fisher:1978pf}, which is most acute in $4 < d < 6$. The equation of state is expected to be mean-field-like in this case, yet, near the Lee-Yang point the critical behavior must be given by nontrivial critical exponents of the $\phi^3$ theory. For $d < 4$ the equation of state must approach the mean-field form as $\epsilon\to0$, yet this cannot be true near the Lee-Yang point because the $\phi^3$ theory is nonperturbative at $d = 4$. We identify and quantify the solution to this apparent paradox. We show that the $\epsilon$ expansion must break down and the equation of state becomes nonperturbative in the (Ginzburg) region around the Lee-Yang point whose radius is proportional to $\epsilon^2$ as $\epsilon\to0$.

We have considered the parametric representation to order $\epsilon^2$ (and $\epsilon^3$, see Appendix) and have shown that the singularities we find are consistent with the Fonseca-Zamolodchikov conjecture (for a range of parameters controlling the form of the parametric representation). However, we have also confirmed that the expansion breaks down near the Lee-Yang edge singularities in a way consistent with the derived Ginzburg criterion. In particular, the order $\epsilon^3$ contribution modifies the results obtained at order $\epsilon^2$ also at order $\epsilon^2$! In other words, the behavior near the singularities (including their position) is nonperturbative at order $\epsilon^2$. Since the distance between the Lee-Yang edge singularity at $t < 0$ (i.e., the spinodal point) from the real axis is itself of order $\beta\delta - 3/2 = \ord(\epsilon^2)$ we conclude that the $\epsilon$ expansion cannot be used to confirm or invalidate the Fonseca-Zamolodchikov conjecture.
 
We point out that the equation of state of the $\O(N)$-symmetric $\phi^4$ theory satisfies the Fonseca-Zamolodchikov conjecture in the large-$N$ limit. In particular, for $d < 4$ there are no singularities on the metastable branch of the real $H$ axis. Instead the singularities can be found off the real axis, and are, in fact, the Lee-Yang branching points, as predicted by extended analyticity. We have checked that (at least in $d = 3$) this result is not affected by the leading $1/N$ corrections.

Although the Fonseca-Zamolodchikov conjecture for the Ising critical equation of state is difficult to prove using the analytic methods considered, we can conclude that it is nontrivially consistent with the various systematic approximations to the equation of state beyond the mean-field level.

The absence of singularities on the real $H$ axis (except for the branch point at $H = 0$ associated with the Langer cut) could have implications for the behavior of systems undergoing cooling past the first-order phase transition (see, e.g., Refs.\ \cite{Randrup:2003mu,Koch:2005pk,Sasaki:2007db,Sasaki:2007qh}). In particular, it could prove important for the understanding of the experimental signatures of the first-order phase transition separating hadron gas and quark-gluon plasma phases of QCD associated with the QCD critical point, which is being searched for using the beam energy scan heavy-ion collision experiments. 

It is important to realize that in the region of the parameter space where the spinodal singularities occur the equation of state is not, strictly speaking, defined in the usual sense as a property of the system {\it in thermal equilibrium}, due to the finite lifetime of the metastable state. It is, however, defined mathematically by analytic continuation from the regime of thermodynamic stability. Many properties of the equation of state in the metastable region, such as the imaginary part and the discontinuity on the Langer cut are clearly reflecting dynamics of the system associated with the decay of the metastable state. Also the absence of the spinodal singularities at real $H$ can be related to metastability: the presence of a thermodynamic singularity requires the correlation length to diverge and the equilibration to such a critical state requires infinite time, which is impossible due to the finite lifetime of the metastable state.

Finally, it is also interesting to note that the decay rate of the metastable state, which is controlled by the (small) coupling $\giv$ (see Eq.\ \eqref{eq:LangerSingularity}) is no longer exponentially suppressed at the spinodal point. Moreover, for small $\givt$, the nucleation rate near the spinodal point has the asymptotic form $\exp[-{\rm const}(w-w_{\rm LY})^{(6-d)/4}/\givt]$ \cite{Patashinskii:1980,Unger:1984}. Therefore exponential suppression disappears in the same region as defined by the Ginzburg criterion in Eq.\ \eqref{eq:w}. This is to be expected since the fluctuations leading to the decay become important in that region. The fact that the shift of the spinodal singularity into the complex $H$-plane is also due to fluctuation contribution to the ``gap'' exponent $\beta\delta$ suggests that the shift is related to metastability.  It would be interesting to establish a more quantitative relation between this phenomenon and the Fonseca-Zamolodchikov conjecture. We defer full development of this connection to future work.
  
\section*{Acknowledgments}

This material is based on work supported by the U.S. Department of Energy, Office of Science, Office of Nuclear Physics under Award Number DE-FG0201ER41195 and within the framework of the Beam Energy Scan Theory (BEST) Topical Collaboration. This work is also funded by the Swiss National Science Foundation.  

\appendix

\section*{Appendix}

\section{Details of the parametric equation of state at order $\epsilon^2$}
\label{sec:Details of the parametric equation of state at order epsilon^2}

For completeness, and for possible future use, in this Appendix we collect the results which are represented schematically in Sec.\ \ref{sec:Singularities in the epsilon expansion}. In particular, they show explicit dependence (or independence) of expansion coefficients on (arbitrary at this order of $\epsilon$) parameters $b_1$ and $b_2$. To obtain the complete expression for $w_n$ in Eq.\ \eqref{eq:w_n_2nd} up to $\mathcal{O}{(\epsilon^2)}$ one needs to expand each quantity in Eq.\ \eqref{eq:F'z-LPM} up to sufficient order. In particular, we need:
\begin{equation}
  \label{eq:h_3_2nd_full}
  h_3 = -\frac{1}{b^2} = -\frac{2}{3} + \frac{4 b_1}{9} \epsilon +
 \frac{1}{9}\left( -\frac{8 b_1^2}{3} + 4 b_2 \right) \epsilon^2 
+ \ord(\epsilon^3),
\end{equation}
and the normalization parameters in Eqs.\ \eqref{eq:wz-theta-LPM}:
\begin{subequations}\label{eq:zwbar_2nd_full}
\begin{align}
  \bar{z} & = \frac{1}{\sqrt{3}}\left[ 1 + \frac{1}{6} \left(4 b_1 - 1\right) \epsilon + \frac{1}{648} \left( 229 + 45 \pi^2 - 360 \lambda - 144 b_1 ( 2 + 3 b_1 ) + 432 b_2 \right)\epsilon^2 + \ord(\epsilon^3) \right] , \\ 
  \bar{w} & = \frac{1}{\sqrt{3}} \left[ 1 + \frac{2b_1}{3} \epsilon + \frac{1}{24} \left( 7 + \pi^2 - 8\lambda - 8 b_1 ( 1 + 2 b_1 ) + 16 b_2 \right)\epsilon^2 + \ord(\epsilon^3) \right]. 
\end{align}
\end{subequations}
Substituting Eqs.\ \eqref{eq:h_3_2nd_full} and \eqref{eq:zwbar_2nd_full} into Eq.\ \eqref{eq:wz-theta-LPM} and using the ansatz for $\theta_n$, Eq.\ \eqref{eq:SeriesAnsatz1}, we obtain
Eq.\ \eqref{eq:w_n_2nd}, where $\mathcal{O}(\epsilon)$ terms cancel and the $\mathcal{O}(\epsilon^2)$ coefficient
\begin{equation}
\begin{aligned}
\label{eq:w_n_2nd_full}
  \omega^{(2)}(c_n, b_1) = \frac{1}{24} \left( 7 + \pi^2 - 8 \lambda - \ln |c_n|^2 - 8 b_1 (1 + 2 b_1) \right) - \frac{3}{8 c_n^2} - \frac{b_1}{c_n} , 
\end{aligned}
\end{equation}
is independent of the parameter $b_2$. 

\section{Parametric equation of state at order $\epsilon^3$}
\label{sec:Parametric equation of state at order epsilon^3}

Here, we consider the {\it extended} linear parametric model, i.e., Eqs.\ \eqref{eq:ParametricModel1} -- \eqref{eq:ParametricModel3}, in order to examine the robustness of our conclusions at $\mathcal{O}(\epsilon^2)$. In particular, we will show how the $\mathcal{O}(\epsilon^2)$ terms in $|w_n|$ are modified by introducing the $\mathcal{O}(\epsilon^3)$ contributions, which again demonstrates the nonperturbative nature of the problem.

In the extended model,
\begin{equation}
  \label{eq:extendedparametric}
  h(\theta) = \bar{h}(\theta + h_3 \theta^3 + h_5 \theta^5 ) ,
\end{equation}
where $\bar{h}$ is an appropriate normalization constant. In contrast to the parametric model of Sec.\ \ref{sec:Parametric equation of state}, the inclusion of a fifth-order contribution in $\theta$ is necessary to match to the equation of state at order $\epsilon^3$ \cite{Wallace:1973,Wallace:1974}. The coefficients $h_3$ and $h_5$ are given by 
\begin{align}
  \label{eq:h35}
  h_3 & = -\frac{1-e\epsilon^3}{b^2} \nonumber \\
  & = -\frac{2}{3} + \frac{4 b_1}{9} \epsilon + \frac{1}{9}\left( -\frac{8 b_1^2}{3} + 4 b_2 \right) \epsilon^2 + \frac{2}{81} \left(8 b_1^3-24 b_1 b_2+18 b_3+27 e\right) \epsilon^3 + \ord(\epsilon^4) , \\
  h_5 & = -\frac{e\epsilon^3}{b^4} = -\frac{4}{9} e \epsilon^3 + \ord(\epsilon^4) ,
\end{align}
with the parameter 
\begin{equation}
  e = \frac{1}{48} \left( 1 + 2 \lambda - 4\zeta(3) -16 b_1^2 \right) ,
\end{equation}
which is negative for all real-valued $b_1$, and 
\begin{equation}
  b^2 = \frac{3}{2} + b_1\epsilon + b_2\epsilon^2 + b_3\epsilon^3 + \ord(\epsilon^4) ,
\end{equation}
expanded in powers of $\epsilon$. The significance of these parameters becomes clear if we factor decompose Eq.\ \eqref{eq:extendedparametric} to the following form
\begin{equation}
  h(\theta) = \bar{h} \theta \! \left[ 1 - (\theta/b)^2 \right] \!\left[ 1 + e \epsilon^3 (\theta/b)^2 \right] ,
\end{equation}
i.e., $b$ and $e$ are related to the zeros of $h$ on the coexistence line ($t < 0$, $H\to 0$). Note, while $\theta = \pm b$ stays finite in the limit $\epsilon\to 0$, $\theta = \pm b/\sqrt{-e \epsilon^3}$ diverges in the same limit.

\begin{figure}[!t]
  \centering
  \includegraphics[width = 0.6\textwidth]{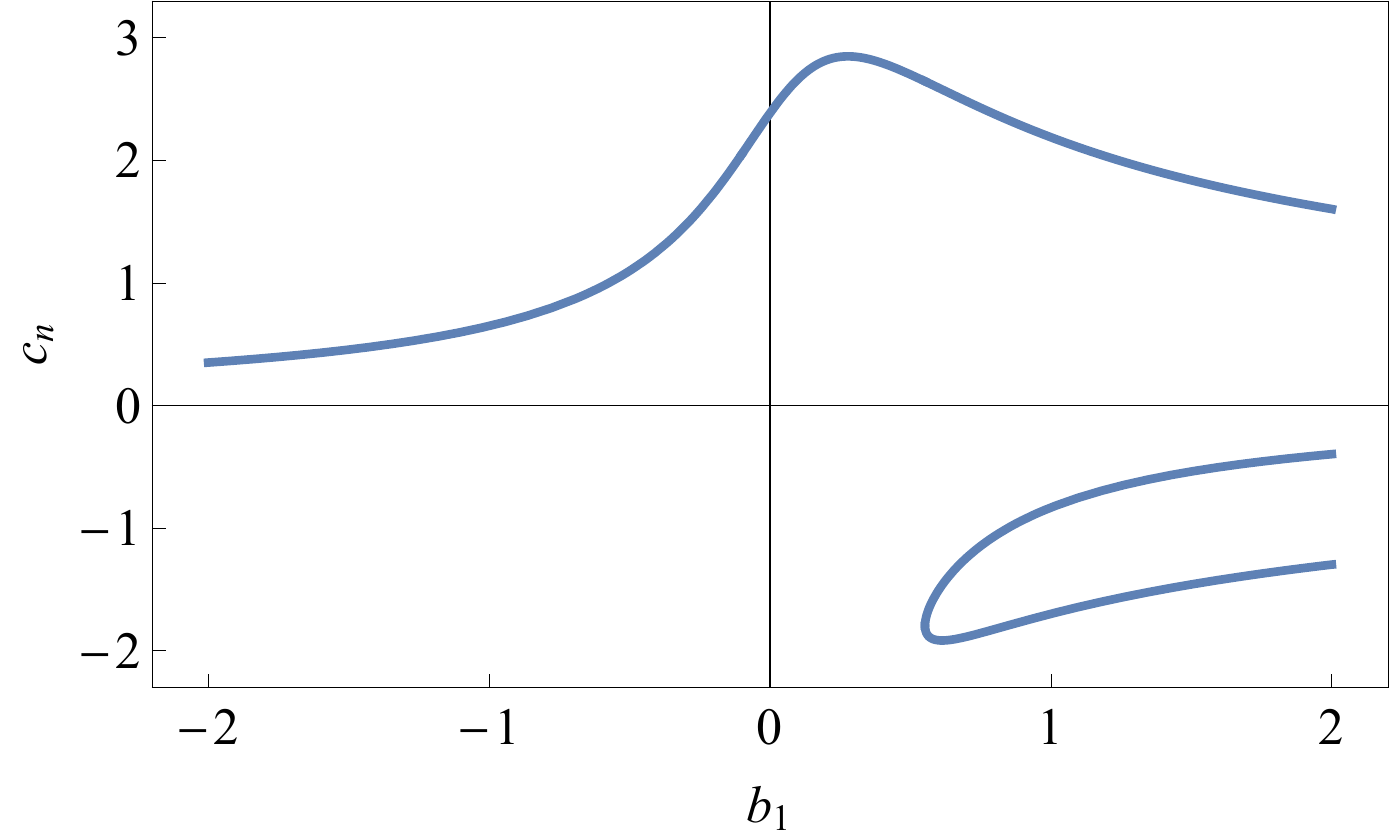}
  \caption{\label{fig:7}The parameters $c_n$, $n = 1,2,3$ as a function of $b_1$. We observe a critical value $b_1\approx 0.552$ above which all solutions $\theta_n^2$, $n = 1,2,3$, are real.}
\end{figure}

To order $\epsilon^3$, the extended linear parametric representation depends on three real-valued parameters $b_1$, $b_2$, and $b_3$. These parameters cannot be fixed by matching to the equation of state alone \cite{Wallace:1974}, which parallels the behavior we have already observed with the order $\epsilon^2$ parametric model (cf.\ Sec.\ \ref{sec:Parametric equation of state}). Essentially, we therefore obtain a three-parameter family of extended linear models that we employ in the following to study the complex-field singularities of the equation of state. 

At order $\epsilon^3$ we find that the (rescaled) inverse susceptibility is given by
\begin{equation}
  \label{eq:F'z-LPM3}
  F'(\theta) = \frac{\bar{w}}{\bar{z}} (1-\theta^2)^{-\gamma}
  \frac{ 1 + \left(2 \beta\delta +3 h_3-1\right) \theta^2 + \left[(2 \beta\delta -3) h_3 + 5 h_5\right] \theta^4 + (2 \beta\delta -5) h_5 \theta^6}{1- (1-2\beta)\hspace{1pt} \theta^2} ,
\end{equation}
where the exponents $\beta$, $\gamma$, and $\delta$, as well as the normalization constants $\bar{w}$ and $\bar{z}$, should be expanded to order $\epsilon^3$ (for details we refer to \cite{Wallace:1973,Wallace:1974}). The zeros of this function, which we consider in terms of $\theta^2$, can be found by solving for the roots of the numerator and can be determined in closed form.

We find three distinct (pairs of) zeros, $\theta_n^2$, which we label by $n = 1,2,3$. It is sufficient to use the ansatz Eq.\ \eqref{eq:SeriesAnsatz1} with the leading-order coefficient given by
\begin{equation}
  \label{eq:SolutionsKappa2} 
  c_n = \frac{1}{24 e} \left[ 1 + \zeta_n + (1 - 288 b_1 e ) \hspace{1pt}\zeta_n^{-1}\right] , \quad n = 1,2,3 , 
\end{equation}
and
\begin{equation}
  \zeta_n = (-1)^{(2/3) (n-1)} \left[1 - 432 b_1 e - 7776 e^2 + \sqrt{\left(1 - 432 b_1 e - 7776 e^2\right)^2 - \left(1 - 288 b_1 e\right)^3} \right]^{1/3} ,
\end{equation}
a function of $b_1$ only. In Fig.\ \ref{fig:7} we illustrate the real-valued coefficients $c_n$ in the range of parameters \mbox{$-2 \leq b_1 \leq 2$}. Note that there is a ``critical'' value of $b_1\approx 0.552$ above which all values of $c_n$ are real -- this has interesting implications as we show below.

\begin{figure}[!t]
  \centering
  \includegraphics[width = 0.3\textwidth]{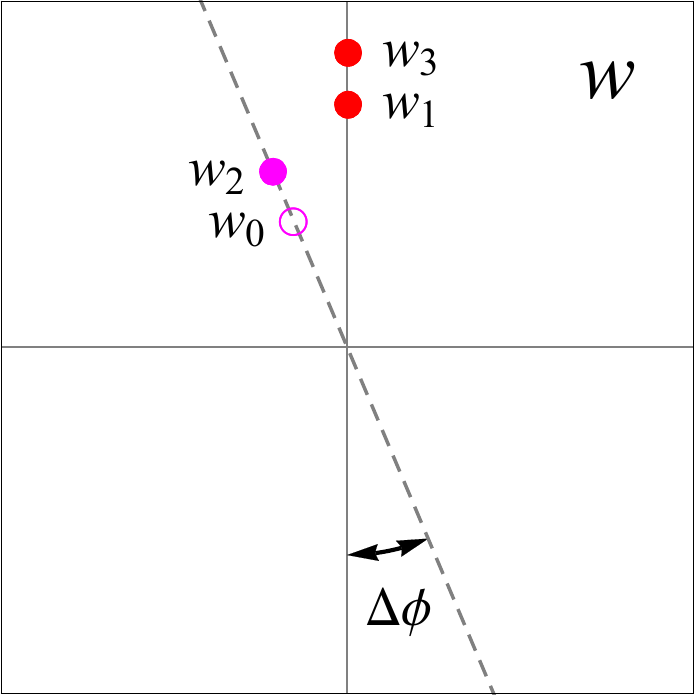}
  \caption{\label{fig:8}We show the distribution of the zeros $w_1$, $w_2$, and $w_3$ (solid points) and pole $w_0$ (open circle) of the parametrized inverse isothermal susceptibility $F'(z)$ up to $\ord(\epsilon^3)$. Here, $b_1 \gtrsim 0.552$, such that all singular points align either along the Lee-Yang cut (situated on the imaginary $w$ axis) or the Langer cut (along the dashed line). Note, only the singularities in the upper half of the complex $w$ plane are shown.}
\end{figure}

As in Appendix. \ref{sec:Details of the parametric equation of state at order epsilon^2}, we finally obtain
\begin{equation}
\begin{aligned}
  w_n =& \pm \frac{2 i \left(-\hat{c}_n\right)^{\frac{3}{2} - \beta\delta}}{3\sqrt{3}}\\
  & \times\: \bigg\{ 1
  + \left[ \frac{1}{24} \left( 7 + \pi^2 - 8\lambda + \ln\frac{\epsilon^2}{c_n^2} - 8 b_1 ( 1 + 2 b_1 ) \right) - \frac{3}{8 c_n^2} - \frac{b_1}{c_n} + \frac{2e c_n}{3} \right] \epsilon^2 + \mathcal{O}(\epsilon^3) \bigg\}, 
\end{aligned}
\end{equation}
where we keep terms up to $\mathcal{O}(\epsilon^2)$ to compare with Eq.\ \eqref{eq:w_n_2nd} and Eq.\ \eqref{eq:w_n_2nd_full}. It is clear that the $\mathcal{O}(\epsilon^2)$ term of the absolute value is modified by the term with parameter $e$, which appears at $\mathcal{O}(\epsilon^3)$ in the extended $h(\theta)$ (see Eq.\ \eqref{eq:h35}). However, the additional corrections to the absolute value do not affect the {\em phases\/} of the corresponding singularities in the complex $w$ plane.

The above results lead to the following picture, which depends on the parameter $b_1$: If $b_1 \gtrsim 0.552$, two points $w_1$ and $w_3$ are imaginary and thus distribute along the Lee-Yang cut, while another point is located on the Langer cut, i.e., $\hat{w}_2 = \pm i (-1)^{3/2 - \beta\delta}$. At $b_1\approx 0.552$ the two zeros $w_1$ and $w_3$ collide and move off the imaginary axis, into the complex $w$ plane, while $w_2$ remains on the Langer cut. On the other hand, from Eq.\ \eqref{eq:F'z-LPM3}, we also obtain a pole, $\theta_0^2$, determined by the same equation as Eq.\ \eqref{eq:Zeros2}, albeit with the critical exponent $\beta$ expanded to order $\epsilon^3$. Thus, this pole is displaced from the imaginary axis and located along the Langer cut, i.e., $\hat{w}_0 = \pm i (-1)^{3/2 - \beta\delta}$ (see Fig.\ \ref{fig:8}).

Summarizing, it appears that the free parameters $b_1$, $b_2$, and $b_3$ can be chosen in such a way that the zeros and poles of the inverse isothermal susceptibility $F'(z)$ align either on the Lee-Yang and/or the Langer cut. If $b_1 \gtrsim 0.552$, there are always singular points located on the Lee-Yang cut, which we might identify with the Lee-Yang edge singularities. This observation supports our earlier suspicion on the nature of rational approximations of functions with a branch cut.

\bibliographystyle{utphys}
\bibliography{references}

\end{document}